\documentclass[12pt]{article}
\usepackage{amsmath}
\usepackage{epsf}
\usepackage{epsfig}

\newcommand{\be}{\begin{equation}}
\newcommand{\ee}{\end{equation}}
\newcommand{\bear}{\begin{eqnarray}}
\newcommand{\ear}{\end{eqnarray}}
\newcommand{\ba}{\begin{array}}
\newcommand{\ea}{\end{array}}

\def\mub{\overline{\mu}}

\makeatletter
\def\gsim{\compoundrel>\over\sim}

\def\compoundrel#1\over#2{\mathpalette\compoundreL{{#1}\over{#2}}}
\def\compoundreL#1#2{\compoundREL#1#2}
\def\compoundREL#1#2\over#3{\mathrel
         {\vcenter{\hbox{$\m@th\buildrel{#1#2}\over{#1#3}$}}}}
\makeatother

\def\l{\left}
\def\r{\right}
\begin{document}

\begin{center}

    {\Large\bf Split NMSSM with electroweak baryogenesis}
    \\
    \vspace{0.8cm}
    \vspace{0.3cm}
    S.~V.~Demidov$^{a,b,}$\footnote{{\bf e-mail}: demidov@ms2.inr.ac.ru}, 
    D.~S.~Gorbunov$^{a,b,}$\footnote{{\bf e-mail}: gorby@ms2.inr.ac.ru},
    D.~V.~Kirpichnikov$^{a,}$\footnote{{\bf e-mail}: kirpich@ms2.inr.ac.ru}
    \\
    
    $^a${\small{\em 
        Institute for Nuclear Research of the Russian Academy of Sciences, }}\\
      {\small{\em
          60th October Anniversary prospect 7a, Moscow 117312, Russia
      }
      }
      \\
$^{b}${\small{\em
Moscow Institute of Physics and Technology,
}}\\
{\small{\em
Institutsky per. 9, 
  Dolgoprudny 141700, Russia
}}\\
  \end{center}
  \begin{abstract}
In light of the Higgs boson discovery we reconsider generation of the
baryon asymmetry in the non-minimal split Supersymmetry model with
 an additional singlet superfield in the Higgs sector. 
We find that
successful baryogenesis during the first order electroweak phase
transition is possible within phenomenologically viable part of
the model parameter space.  We discuss several phenomenological consequences of this
scenario, namely, predictions for the  electric dipole moments of
electron and neutron and collider signatures of light charginos
and neutralinos.
  \end{abstract}

\section{Introduction}

Any phenomenologically viable particle physics model should explain 
the observed asymmetry  between  matter and antimatter in the
Universe. The analysis of the anisotropy and polarization of the cosmic microwave background provided by WMAP
collaboration gives the following baryon-to-photon
ratio~\cite{Bennett:2012zja} 
\be 
 \frac{n_B}{n_\gamma} = 
(6.19 \pm 0.14) \times 10^{-10}. 
\label{BaryonAsymm1}
\ee 
To generate the baryon asymmetry of the Universe, three Sakharov's
conditions should be satisfied~\cite{Sakharov:1967dj}: (i) baryon
number violation, (ii) $C$- and $CP$-violation and (iii)  departure from
thermal  equilibrium. The latter condition can be realized, in
particular, during the strong first order electroweak phase transition
(EWPT) which proceeds via nucleation and expansion of bubbles of new
phase in the hot plasma of  the early Universe (for a recent discussion
see, e.g., Refs.~\cite{Morrissey:2012db,Konstandin:2013caa}). The
baryon number violation during  the EWPT happens due to sphaleron processes
in symmetric phase, while the $CP$-violation is induced by the interaction
of particles in plasma with the  bubble walls. 
  
In the  Standard Model of particle physics (SM) the Sakharov's
conditions are only partly fulfilled. In particular, baryon number is
violated via electroweak sphaleron transitions at high
temperatures. At the same time, the electroweak transition in the SM
is not the first order phase transition, hence no 
sufficient departure from thermal equilibrium. And the contribution of
$CP$-violating CKM phases is too small in any case to  provide
(\ref{BaryonAsymm1}). Finally, the electroweak sphalerons in the
broken phase are too fast and would wash out any baryon asymmetry
generated during  the EWPT~\cite{Kajantie:1996mn,Csikor:1998eu}. 
Therefore, electroweak baryogenesis is only possible in SM
extensions.  These models should contain additional sources of 
$CP$-violation. Moreover, if the baryon asymmetry emerges at the
electroweak scale, there should be a mechanism making the EWPT to
be the strongly first order. A lot of scenarios for baryogenesis during
the EWPT have been proposed and
studied,  see e.g.
Refs.~\cite{Bodeker:2004ws,Fromme:2006cm,Li:2008ez,Kozaczuk:2012xv,Huber:2000mg,Huber:2000ih,Huber:2006wf,Kozaczuk:2013fga,Cheung:2012pg}.    
%

The  Minimal Supersymmetric Standard Model (MSSM) is one of the most
elegant ways to extend the SM framework.  In particular, the quadratic
divergences cancellation and the gauge couplings unification are
the major reasons for the interest in supersymmetric 
models. Moreover, the  
lightest neutralino  is a natural dark matter candidate in the MSSM~\cite{Feng:2010gw,Martin:1997ns}.
In general, however, the Higgs boson
discovery~\cite{Chatrchyan:2012xdj,Aad:2012tfa}, and non-observation
of superpartners at the LHC shrinks severely the region
of MSSM parameter space. For instance, squarks and gluinos have been
searched for at the LHC~\cite{Aad:2012fqa,Chatrchyan:2012jx}, and the
lower bounds on their masses have been set at the level of 1-2 TeV.  
  
An attractive  MSSM extension with splitted superpartner spectrum  
(split MSSM) has been
proposed in Refs.\,\cite{ArkaniHamed:2004fb,Giudice:2004tc}. The squarks and 
sleptons in these scenarios are very heavy, while neutralinos and
charginos remain light. Nevertheless, the main advantages of SUSY,
i.e. the gauge coupling unification and existence of dark matter
candidate,  remain intact in this class of models. Remarkably,
the absence of FCNC processes\,\cite{Agashe:2014kda} is 
naturally understood within this setup. Unfortunately, the electroweak
baryogenesis can not be realized in minimal version of the split
SUSY. This can be cured by introducing a gauge singlet superfield to 
the Higgs sector  of the split MSSM~~\cite{Demidov:2006zz}. 
The main features of this split Next-to-Minimal Supersymmetric
extension of the Standard Model, {\it split NMSSM}, are the following.
There are two energy scales in the split NMSSM, electroweak $M_{EW}\sim100$\,GeV
and splitting scale $M_S\gg M_{EW}$. At $M_{EW}$ scale, the spectrum of split
NMSSM contains the SM particles, one Higgs doublet $H$, the higgsino
components  $\tilde{H}_{u,d}$,  winos $\tilde{W}$,  bino $\tilde{B}$,
and in addition a singlet  complex scalar field $N$ and its superpatner
singlino $\tilde{n}$. The sleptons, squarks and four out of seven
scalar degrees of freedom in the Higgs sector have masses of order
the splitting scale $M_S$. Hence, these particles are decoupled from
the spectrum at low energies $E<M_{S}$. At the same time, 
  interactions of the scalar components of the singlet $N$ with the
  Higgs  boson are 
described at $M_{EW}$ by a generic potential, which includes trilinear
terms.  These couplings are capable of
strengthening the first order EWPT.  In the present paper, we
review this scenario in view of the latest experimental results, in
particular, the Higgs boson discovery. 

This paper is organized as follows. In Section~\ref{SplitNMSSM} we
discuss the structure of split NMSSM.  In Section~\ref{SectionScan},
we explore the phenomenologically allowed region of the model
parameters consistent  with the Higgs boson of mass $m_H~\simeq
  125$\,GeV.  In  
Sections~\ref{SectPhaseTrans}  and
\ref{BaryAsymmSect} we study the strong first order 
EWPT and
 the baryon asymmetry of the Universe, respectively,  for the 
relevant split
NMSSM  parameter space. In Section~\ref{EDMLHC} we perform an 
analysis
of the electron and neutron EDMs. There we  also
discuss  the spectra of charginos and neutralinos, which can be 
probed at the  LHC experiments. In
Appendix~\ref{Appendix1LoopHiggsMass} we calculate one-loop
renormalization group (RG) corrections to the Higgs boson mass, which 
are needed to find allowed region of the parameter space in the  
split NMSSM scenario. In Appendix~\ref{SectFineTunDim} the 
minimization conditions for the split NMSSM effective 
potential are presented.

\section{ Non-minimal split Supersymmetry
\label{SplitNMSSM}} 

In this Section we discuss the Lagrangian  and particle content of the
split NMSSM. Above the splitting scale $M_S$, the model is
described by  generic\footnote{A quadratic in $\hat{N}$ term can be
  eliminated by a field redefinition.} NMSSM superpotential 
\be 
 W=\lambda \hat{N} \hat{H}_u \epsilon \hat{H}_d+
 \frac{1}{3} k \hat{N}^3+\mu \hat{H}_u\epsilon \hat{H}_d + r \hat{N},
 \label{SupPotNMSSM}
\ee 
where  $\hat{H}_{u,d}$ are superfields of the Higgs doublets,
$\hat{N}$ is a chiral superfield singlet with respect to
$SU(3)_C\times SU(2)_L\times U(1)_Y$ gauge group, $\hat{N}=N+\sqrt{2}\,\theta
\tilde{n}+\theta^2 F_N$ , and $\epsilon$ is antisymmetric
$2\times 2$ matrix with $\epsilon_{12}=-\epsilon_{21}=1$. 

The tree level scalar potential of the non-minimal SUSY model can be
written  as follows
\be 
V=V_D+V_F+V_{soft},
\label{VdVfVs}
\ee
where the contribution of $D$-terms is the same as that in the MSSM, 
$$
V_D=\frac{g^2}{8}
\left(H_d^\dagger \sigma_a H_d + H_u^\dagger \sigma_a H_u\right)^2+ 
\frac{g'^2}{8}\left(|H_d|^2-|H_u|^2\right)^2,
$$
with $g$ and $g'$ being $SU(2)_L$ and $U(1)_Y$ gauge 
couplings, respectively.
The contribution of $F$-terms derived from superpotential 
(\ref{SupPotNMSSM}) reads
$$
V_F=\left|\lambda H_u \epsilon H_d +k N^2 +r\right|^2
+|\lambda N +\mu|^2 \left(H_u^\dagger H_u+H_d^\dagger H_d\right).
$$
Soft supersymmetry breaking terms are described by the potential
\begin{gather}
\label{bpotential}
  V_{soft} = \l( \lambda A_\lambda N H_u \epsilon H_d +
\frac{1}{3}k A_k N^3 + \mu B H_u \epsilon H_d +A_r N +\text{h.c.}\r) \\
+m_u^2 H_u^\dagger H_u + m_d^2 H_d^\dagger H_d 
+ m_N^2 |N|^2, \label{VSoft}
\end{gather}
where  $A_{\lambda, k}$ and $m_{u,d,N}$ are the trilinear couplings
and the soft masses of scalars, respectively. Components of the Higgs
doublets $H_{u,d}$ and singlet field $N$  
in \eqref{bpotential}, (\ref{VSoft}) are defined
by
\be
 H_u =
\left(
\begin{array}{c}
H_u^+\\
 H_u^0
\end{array}
\right), \quad 
 H_d =
\left(
\begin{array}{c}
H^0_d\\
 H^-_d
\end{array}
\right), \qquad  N  = (S+iP)/\sqrt{2},
\label{HiggsSinglet}
\ee 
where $S$ and $P$ are the scalar and pseudoscalar parts of the singlet
$N$, correspondingly.  We introduce the following notations:  
$
\tan \beta \equiv \langle H^0_u \rangle / \langle H^0_d \rangle
$, $v_S\equiv\langle S\rangle$ and $v_P\equiv\langle P\rangle$.

 An explicit analysis of the particle spectrum  of the model with the
  potential
(\ref{VdVfVs}) is performed in Ref.~\cite{Demidov:2006zz}. We
nevertheless briefly discuss the particle content of
 the  scalar sector 
at energies below the splitting scale. There are ten scalar degrees of 
freedom  at the splitting scale $M_S$, coming  from (\ref{HiggsSinglet}).
It is shown in Ref.~\cite{Demidov:2006zz} that if the soft SUSY
breaking parameters $B \mu$, $m_d^2$ and $m_u^2$ are of order of the 
squared splitting scale, $M^2_S$, then two charged Higgses,
 one pseudoscalar and one neutral scalar Higgs bosons are heavy
and thus decoupled from the low energy spectrum, while a fine-tuning
is required for the mass of the lightest Higgs boson $H$ and two
singlets, $S,P$ to be at the electroweak scale. Three Goldstone modes
are  eaten by $W^\pm$ and $Z^0$ due to the Higgs mechanism. 
We emphasize that the particle spectrum in the split NMSSM (as
well as in any split SUSY model) below $M_S$ requires a  fine-tuning of
the soft dimensionful parameters~\cite{Demidov:2006zz}.

 Replacing $H_u \rightarrow H \sin \beta$ and $H_d \rightarrow
\epsilon H^* \cos \beta$ in (\ref{VdVfVs}) we obtain at the splitting
scale $M_S$ the effective
Lagrangian for the relevant at low energy degrees of freedom in the
scalar sector  of
the model 
 (hereafter we omit the corresponding kinetic terms),  
$$
-\mathcal{L}_V=\frac{\bar{g}^2}{8}\cos^2 2\beta 
(H^\dagger H)^2+
|r+kN^2-\frac{\lambda}{2}\sin 2\beta H^\dagger H|^2+
|\lambda N+\mu|^2 H^\dagger H
$$
\be 
+\left(-\frac{\lambda}{2} A_\lambda \sin 2 \beta N H^\dagger H
-\frac{\mu B}{2} \sin 2 \beta H^\dagger H+
\frac{1}{3}k A_k N^3 +A_r N +h.c.
\right)
\label{LScalatMs}
\ee 
$$
+(m_u^2 \sin^2 \beta +m_d^2 \cos \beta )H^\dagger H
+m_N^2 |N|^2,
$$
where $\bar{g}^2\equiv g^2+(g')^2$.  The quark-Higgs Yukawa
interactions, gaugino couplings and gaugino mass terms are the same as
in the minimal split supersymmetry. New part of the Yukawa
interactions for  Higgsinos  $\tilde{H}_{u,d}$ and singlino field 
$\tilde{n}$ is given by  
\be 
-\mathcal{L}_Y=-\lambda N \tilde{H}_u \epsilon \tilde{H}_d
-\lambda \sin\beta H^T \epsilon (\tilde{H}_d \tilde{n})
+\lambda \cos \beta (\tilde{n} \tilde{H}_u) H^*
-k N \tilde{n} \tilde{n}+h.c.
\ee 
Now we consider the most general  scalar Lagrangian at 
energies below $M_S$  
$$
-{\cal L_{\rm{V}}} = -m^{2}H^{\dagger}H + 
\frac{\tilde{\lambda}}{2}\l(H^{\dagger}H\r)^{2} + 
i\tilde{A}_{1}H^{\dagger}H\l(N^* - N\r) + 
\tilde{A}_{2}H^{\dagger}H\l(N + N^{*}\r) +
2 \kappa_{1}|N|^{2}H^{\dagger}H 
$$
\be
  \label{gener_poten}
 + \kappa_{2}H^{\dagger}H\l(N^{2} + N^{*2}\r) + 
  \tilde{m}_{N}^{2}|N|^{2} + \lambda_{N}|N^{2}|^{2} + 
  \frac{1}{3}\tilde{A}_{k}\l(N^{3} + N^{*3}\r)
+ \tilde{A}_{r}\l(N + N^{*}\r) 
\ee
$$
+ \l(\frac{\tilde{m}^{2}}{2}N^{2} + \frac{1}{2}\tilde{A}_{3}N^{2}N^{*} +
\xi N^{4} + \frac{\eta}{6}N^{3}N^{*} + h.c.\r),
$$
here  the quartic couplings $\tilde{\lambda}$, $\kappa$, 
$\kappa_1$, $\kappa_2$ and $\lambda_N$ at  the electroweak scale
are related via renormalization group equations to $\bar{g}$,
$\lambda$, $k$ and $\tan \beta$ at the scale  
$M_S$.  Comparing scalar potential (\ref{gener_poten}) with
(\ref{LScalatMs}) one can obtain the matching conditions for these 
couplings at the splitting scale $M_S$:
\be 
\kappa_1=\lambda^2, 
\quad \kappa_2=-\lambda k \sin \beta \cos \beta,
\quad \lambda_N =k^2, \quad \kappa= \lambda,  
\label{a1}
\ee
\be
\tilde{\lambda}=\frac{\bar{g}^2}{4}\cos^2 2\beta +
\frac{\lambda^2}{2} \sin^2 2 \beta. \label{a666} 
\ee
We use here the convention $g_1^2=(5/3)g'^{2}$ and $g_2=g$
adopted in Grand Unified Theories (GUT).    
Note that  the couplings proportional to $\xi$ and $\eta$ in
(\ref{gener_poten}) are absent in the effective Lagrangian at
$M_S$, but get induced  by loop quantum corrections; thus we
set the following RG initial condition  
\be 
\xi=\eta=0
\label{xietazero}
\ee  
at the splitting scale $M_S$.  Soft fermion masses and Yukawa
interactions below $M_S$ are described by the Lagrangian 
$$
- {\cal L_{\rm{Y}}} = \frac{M_{2}}{2}\tilde{W}^{a}\tilde{W}^{a} + 
\frac{M_{1}}{2}\tilde{B}\tilde{B} +
\l(\mu + \kappa N\r)\tilde{H}^{T}_{u}\epsilon\tilde{H}_{d} -
kN\tilde{n}\tilde{n} 
$$
\be
+ H^{\dagger}\l(\frac{1}{\sqrt{2}}\tilde{g}_{u}\sigma^{a}\tilde{W}^{a}
+ \frac{1}{\sqrt{2}}\tilde{g}_{u}^{\prime}\tilde{B} -
\lambda_{u}\tilde{n}\r)\tilde{H}_{u}   \label{gener_yukava} 
\ee
$$
+ H^{T}\epsilon\l( 
-\frac{1}{\sqrt{2}}\tilde{g}_{d}\sigma^{a}\tilde{W}^{a} +
\frac{1}{\sqrt{2}}\tilde{g}^{\prime}_{d}\tilde{B} - 
\lambda_{d}\tilde{n}\r)\tilde{H}_{d} + h.c.,
$$ 
where $M_2$ and $M_1$ are wino and bino soft mass parameters in $SU(2)_L$
and $U(1)_Y$ gaugino sectors, respectively. The corresponding matching
conditions for Yukawa couplings at the splitting scale $M_S$ read 
\be  
\qquad \lambda_u=\lambda \cos \beta,
\qquad \lambda_d=-\lambda \sin \beta,
\label{a2}
\ee
\be
\tilde{g}_u=g \sin \beta, \qquad \tilde{g}_d=g \cos \beta, \qquad
\tilde{g}'_u=g' \sin \beta, \qquad \tilde{g}'_d=g' \cos \beta.
\label{a3}
\ee
Matching equations for the dimensionful couplings in
(\ref{gener_poten}) can be found in a similar way. However, for
simplicity we take their values directly at  electroweak scale rather
than solving RG equations for them from $M_S$ down to electroweak
energies. In order to reduce the number of trilinear couplings we
 assume that Higgs-scalar $(H-S)$ and Higgs-pseudoscalar $(H-P)$ 
 mixing terms in their squared mass matrix are equal to zero at
  the EW energy scale. This implies appropriate  relations
  for the  trilinear couplings $\tilde{A}_1$ and $\tilde{A}_2$, 
\be
\tilde{A}_1= \sqrt{2}(\kappa_1-\kappa_2)v_P, \qquad 
\tilde{A}_2= - \sqrt{2}(\kappa_1+\kappa_2)v_S.
\label{A1A2} 
\ee 
 From the very beginning we admit explicit $CP$-violation by taking
purely imaginary $\mu$-term and from largangians (\ref{gener_poten})
and (\ref{LScalatMs})  we relate its value through the following
matching condition at 
$M_S$  scale   
\be 
\text{Im}\, \mu =\tilde{A}_1/\lambda
\label{A1muDef}
\ee
neglecting small RG corrections.
 Let us note that using minimization conditions for the
potential~(\ref{gener_poten}), soft squared masses $m^2$, 
$\tilde{m}^2$ and $\tilde{m}^2_N$ can be re-expressed via vevs of the
scalar fields, i.e. $v$, $v_S$  and $v_P$. For completeness these
relations are presented in Appendix\,\ref{SectFineTunDim}. 

 With the all above assumptions, we are left with only seven
independent dimensionful parameters of the model at the EW scale
\be(v_S, \quad v_P, \quad 
M_1, \quad  M_2, \quad \tilde{A}_k, \quad  \tilde{A}_3,
\quad \tilde{A}_r).
\label{FreeSoftTerms}
\ee
In what follows to get numerical results, for concreteness, we set at
the EW scale:    
\be
 \tilde{A}_3 = \tilde{A}_r = 0, \quad \tilde{A}_k= -1.1 \text{ GeV,}
\ee
while scanning over all the other four parameters. We advertise that
the two singlet VEVs $v_S$ and  $v_P$  play very prominent role 
in developing the EWPT,  which is discussed below in Section~\ref{SectPhaseTrans}.

\section{Predictions for the Higgs boson mass
\label{SectionScan}}
 
In this Section  we describe  the scanning over the set of three
dimensionless parameters $(\tan \beta, \lambda, k)$ fixed at scale
$M_S$  and calculate the mass of the Higgs
  boson resonance. We outline the region of model parameter
space  consistent with the SM-like Higgs boson with mass about
  125\,GeV.  

In our procedure we choose dimensionless couplings of the model
at the splitting scale and calculate the value of the Higgs boson
mass by solving RG equations at next-to-leading order in coupling
constants (NLO). 
We start solving the truncated  part of the RG equations from the
EW up to the splitting scale for the SM couplings 
\be 
(g', \quad g, \quad g_s, \quad y_t),
\label{smCoupling}
\ee 
where $g_s$ is $SU(3)_c$ gauge coupling and $y_t$ denotes the top
Yukawa coupling. Initial conditions for RG equations for these
couplings at the EW scale are taken as
  follows~\cite{Agashe:2014kda}
 \[
\alpha_s(M_Z) =0.118\,, \quad 
 M_Z=91.19\,\text{GeV}\,, \quad M_W=80.39\, \text{GeV, \,\,and} \quad
 y_t(m_t)=0.95.
\] 
Next, we use  complete set of the RG  equations for dimensionless 
couplings of the split NMSSM  
\be
(g',\quad g,\quad g_s,   \quad  y_t, \quad 
\tilde{\lambda}), \quad \quad (\tilde{g}_{u,d}, \quad \tilde{g}'_{u,d}, \quad  
 \lambda_{u,d},  \quad \kappa, \quad
\kappa_{1,2}, \quad k, \quad \lambda_N,
\quad \xi, \quad \eta).
\label{splitCouplings}
\ee 
Corresponding RG equations can be found
in~Ref.~\cite{Demidov:2006zz}. In order to obtain values of the
couplings~(\ref{splitCouplings}) at low energies, the
values of  $\tan \beta$, $\lambda$ and $k$  are chosen 
randomly at the   splitting scale $M_S$ from the following
perturbative regions 
\be
-0.6<k<0.6, \quad  0<\lambda<0.7, \quad 0<\tan\beta< 30.
\label{ranges}
\ee 
 Then we solve the complete set of the RG equations from $M_S$
down to the EW scale by using matching condition (\ref{a1}), 
(\ref{xietazero}), (\ref{a2}) and (\ref{a3}). 
 This procedure  doesn't guarantee the correct value of top Yukawa
  coupling at low energy $y_t(M_t)$. Therefore, we  tune $y_t(M_S)$
  to obtain the value of $y_t(M_t)$ within the error bars  
 (for details see \ref{TopYukawaAppendix} and
  Refs.~\cite{Demidov:2006zz,Binger:2004nn}).
 We include a part of 
threshold correction to the Higgs quartic coupling at the splitting
scale~~\cite{Giudice:2011cg} resulted in the following 
modification  
\be 
\tilde{\lambda}\rightarrow \tilde{\lambda}+\delta \tilde{\lambda}, 
\ee
where $ \delta \tilde{\lambda} $ is a conversion term  from 
$\overline{\mbox{DR}}$ to $\overline{\mbox{MS}}$ renormalization
schemes  at $M_S$,  
\be 
\delta \tilde{\lambda}=-\frac{1}{16\pi^2}\Bigl[ \frac{9}{100}g_1^4+
\frac{3}{10}g_1^2 g_2^2+\left(\frac{3}{4} - \frac{\cos^2
  2\beta}{6}\right)g_2^4 +
\frac{3}{400}(5g_2^2+3g_1^2)^2\sin^24\beta\Bigr].  
\ee
 The remaining part of the threshold correction to $\tilde{\lambda}$
  depends on hierarchy of masses of heavy scalars near the splitting
  scale and it has not been taken into account. We should keep it in
  mind when interpreting the results.
Next, we calculate the pole mass of the Higgs boson including
 one-loop threshold corrections at the electroweak scale, see
Appendix~\ref{Appendix1LoopHiggsMass} for details. 
\begin{figure}[!htb]
\begin{center}
\includegraphics[width=0.8\textwidth]{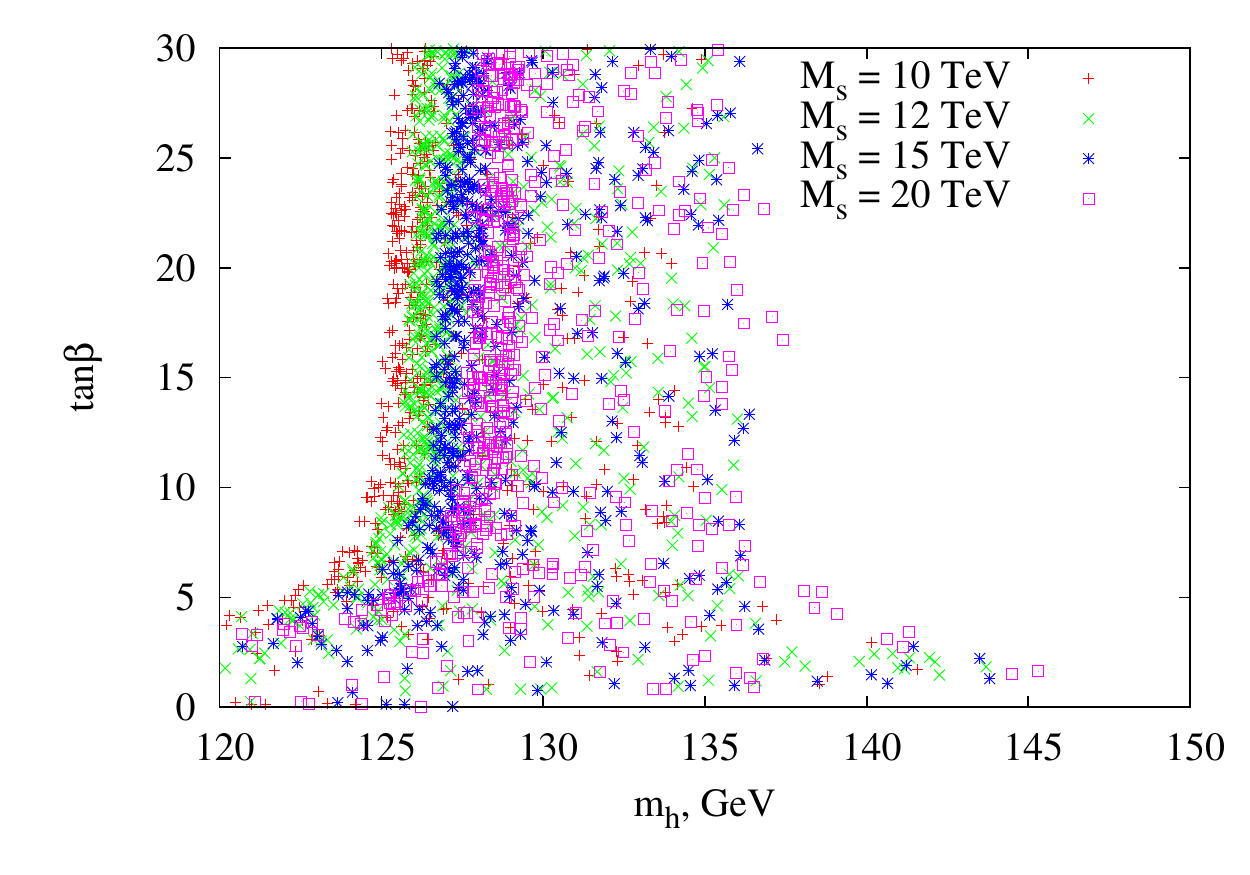}
\caption {Prediction for the Higgs boson mass $m_h$ as a function of
  $M_S$ and $\tan \beta$. We assumed here that the Yukawa top coupling
  falls within the range $y_t^{lower}<y_t<y_t^{upper}$, see the main
  text for details.\label{TgBVsMh}}
\end{center}
\end{figure}  
In Fig.~\ref{TgBVsMh} we show prediction for the Higgs boson mass 
obtained with 
 various values of split scale $M_S$ and $\tan \beta$. It follows
from Fig.~\ref{TgBVsMh} that  for most of the models the Higgs
mass shifts by several GeVs  if one increases  the splitting
  scale $M_S$ from 10 to 20 TeV for $\tan \beta >10$. 
The similar behavior was observed in split
MSSM~\cite{Giudice:2011cg}.  This is   attributed to a large quantum
  correction coming from heavy stops. 

\begin{figure}[!htb]
\begin{center}
\includegraphics[width=0.8\textwidth]{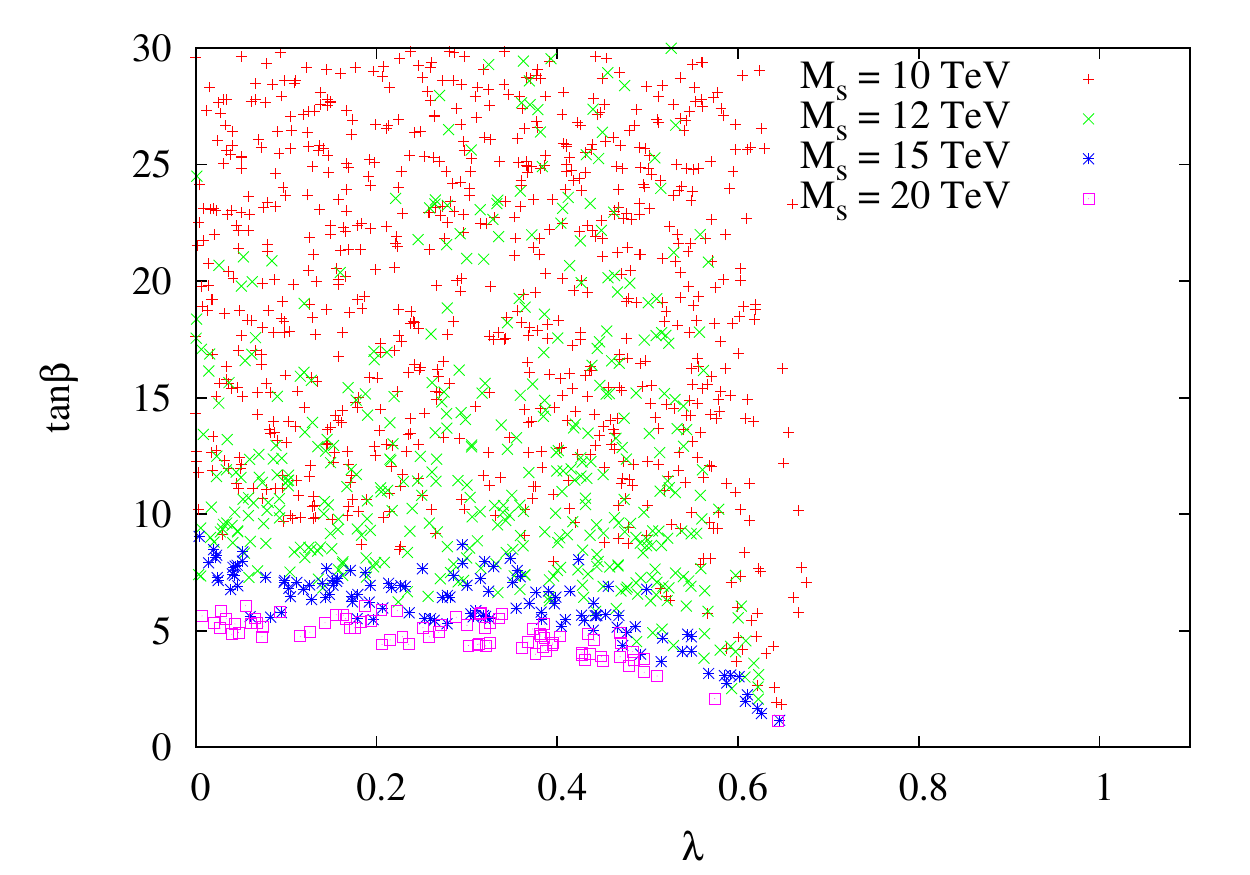}
\caption {Allowed regions for $\tan \beta$ and $\lambda(M_S)$ 
for various values of the splitting scale~$M_S$. 
 \label{figure1}}
\end{center}
\end{figure}

 Now, we require that the  pole mass of the Higgs boson 
(\ref{OneLoopHiggsMass1}) and $y_t$  at $\mub = M_t$ fall  within the 
 following ranges
$$
125.3 \; \mbox{GeV} <m_h^{pole}<125.9 \; \mbox{GeV}, \;\;\;
y_t^{lower}<y_t<y_t^{upper}.
$$
Here we use the average value $m_h=125.6\pm 0.3$\,GeV from
CMS~\cite{Chatrchyan:2012xdj} and ATLAS~\cite{Aad:2012tfa} combined
results (for details see, e.g., Ref.\,\cite{Agashe:2014kda} and
references therein). Lower and upper limits for $y_t$ are extracted
from Eq.~(\ref{Yukawa1}) and correspond to $M_t^{lower}=172.3$\,GeV and
$M_t^{upper}=174.1$\,GeV respectively.  
In Fig.\,\ref{figure1} 
we show  the selected models in $(\tan \beta,\lambda)$-plane for the
 values of the splitting scale $M_S$ varying from 10 to
20~TeV. One can see that for $\tan \beta > 5$ parameter  $\lambda$
  can take arbitrary values in the allowed perturbative region. For
$\tan \beta \simeq 1$ the allowed  region shifts to the maximal 
values of $\lambda$ which follows from the matching condition~\eqref{a666}. 
 We check that $\lambda$ is in the perturbative regime up to the GUT
 scale. 
In addition, as  follows from Fig.\,\ref{figure1}, the 
  phenomenologically possible values of $\tan \beta$ grow with
decreasing of the splitting scale $M_S$ for $\lambda < 0.4$. 
 This is again related to the balance between the tree-level and
  loop-induced contributions to the Higgs boson mass.
  The regions where  $\tan \beta $ is either large $(\beta \rightarrow \pi/2 )$  or small $(\beta \rightarrow 0 )$ correspond to the 
  decoupling of the second term in  (\ref{a666}). 
We find that for $M_S \rightarrow \infty$  the allowed regions for
$\tan \beta$ and $\lambda$ shrink to $\tan \beta \rightarrow 1$ and
$\lambda \rightarrow 0$, respectively. 

  As it follows
from~(\ref{a1}) and (\ref{a666}) the coupling $k$ does not enter the matching condition
for $\tilde{\lambda}$ at $M_S$ 
and we find that the value of the Higgs boson  mass in the model is
almost independent of the coupling constant  $k$ within the
perturbative ranges~(\ref{ranges}). 

In what follows, we choose two  close benchmark setups for the
free parameters  
\be 
\mbox{Setup}\, 1: \qquad M_S =12\,\, \mbox{TeV}\,, \,\, \tan \beta =
9.21\,, \,\, \lambda = 0.559\,,\,\, k=-0.5\,;
 \label{benchsplit1}
\ee
\be 
\mbox{Setup}\, 2: \qquad M_S =10\,\, \mbox{TeV}\,, \,\, \tan \beta = 10.0\,,
\,\, \lambda = 0.611\,,\,\, k=-0.5\,. 
 \label{benchsplit}
\ee
The both benchmark models are well inside the allowed regions in 
Fig.~\ref{figure1}.  For calculation of the threshold correction
  the relevant dimensionful parameters are taken  to be $M_2=
  1$~TeV,   $M_1=300$~GeV and $\mbox{Im}\, \mu = 120$~GeV.   
 As it has been found
  in~\cite{Demidov:2006zz} the resulting 
  baryon asymmetry is directly related to the value of $\lambda$. Thus
  the coupling $\lambda$ is rather large for both chosen  models.
The relevant Yukawa and quartic couplings at the electroweak scale,
$\mub=M_t=173.2$ GeV,  are presented in Table\,\ref{dimlsscplngs}. 
    \begin{table}[t]
\begin{center}
\begin{tabular}{|c|c|c|c|c|c|c|c|c|c|c|}
\hline
 & $ \tilde{g}_u $ & $ \tilde{g}_d $ & $ \tilde{g}_u' $ & $ \tilde{g}_d' $ & $\lambda_u$ &
$ \lambda_d $ & $  \kappa $ & $\kappa_1$ & $ \kappa_2 $ &   $ \lambda_N$ \\
\hline
Setup\,1& $ 0.650 $ & $ 0.070 $ & $ 0.347 $ & $ 0.037 $ & $ 0.057 $ &
$ -0.513 $ & $  0.560 $ & $ 0.251$ & $ -0.022 $  &  
$ 0.208 $ \\
\hline
Setup\,2& $ 0.649 $ & $ 0.065 $ & $ 0.347 $ & $ 0.034 $ & $ 0.056 $ &
$ -0.560 $ & $  0.609 $ & $ 0.297$ & $ -0.021 $  &  
$ 0.207 $ \\
\hline
\end{tabular}
\caption{\label{dimlsscplngs}
 Dimensionless couplings at the electroweak scale. 
}
\end{center}
\end{table}  
Below we use these couplings in the analysis of the strong
first order  EWPT (Section\,\ref{SectPhaseTrans}), in the
  calculation of BAU (Section\,\ref{BaryAsymmSect}) and to estimate
  the values of EDMs of the electron and neutron
  (Section\,\ref{EDMLHC}).

\section{Strong first order EWPT
\label{SectPhaseTrans}}    
In this Section we revisit  the results of
Ref.~\cite{Demidov:2006zz} for the strongly first order electroweak
phase transition in  the split NMSSM within the region of the
parameter space favored by the measured  value of the Higgs boson
  mass ($m_h\simeq 125$ GeV).  
Let us  consider the effective potential at finite temperature $T$
\cite{Dolan:1973qd} 
\be 
V^{eff}_{T}=V_{T=0}^{eff}+V^{(1)}_{T}.
\label{eff_pot1}
\ee 
Here $V^{(1)}_{T}$ is the thermal contribution  
given by 
\be
V_T^{(1)}=\sum_i f_i(m_i,T)\;, 
\label{effpotT1} 
\ee
where sum goes over all  species in the hot plasma (see, e.g., 
Eq.\,(\ref{1l_eff_pot})), and 
\be 
f_i(m_i,T) = (\pm) \frac{T^4}{2\pi^2} \int_0^{\infty} 
dx\, x^2 \ln \left(1 \mp e^{-\sqrt{x^2+(m_i/T)^2}}\right)\;,
\ee
where the upper and lower signs correspond to 
bosons and  fermions respectively. 
 
In order to avoid  baryon number washout after the 
phase transition
the condition $v_c/T_c \gsim 1.1$  has to be satisfied 
\cite{Moore:1998swa}  (see also recent revised discussion
in~\cite{Patel:2011th}). Here $v_c$ is the Higgs VEV at the critical
temperature $T_c$. We define $T_c$ as a temperature at which one
bubble of the broken phase begins to nucleate within a causal
space-time volume of the  Universe. The latter is  determined by
the Hubble parameter  $\mathcal{H}(T)$ as   
\be 
\mathcal{H}^{-4}(T) = (M_{Pl}^*/T^2)^4.
\ee
The bubble nucleation rate in a unit space-volume has the form 
\be
\Gamma(T) \simeq (\text{prefactor})\times T^4 \exp(-S_3/T)\,, 
\ee
where $S_3=S_3(T)$ is the free energy of the critical bubble at
a given temperature 
\be 
S_3(T)= 4\pi \int_0^{\infty} dr \, r^2 
\left[ \frac{1}{2}\left(\frac{dh}{dr}\right)^2+ 
\frac{1}{2}\left(\frac{dS}{dr}\right)^2+ 
\frac{1}{2}\left(\frac{dP}{dr}\right)^2+V^{eff}_T(h,S,P)\right].
\label{FreeEn1}
\ee  
Here $h(r), S(r)$ and $P(r)$ are the radial configurations of the
scalar fields, which  minimize the functional $S_3$. Therefore, the probability that
the bubble is nucleated inside a causal volume reads
\be 
P\sim\Gamma \cdot\mathcal{H}^{-4}\sim \frac{M_{Pl}^{*4}}{T^4}\exp(-S_3/T).
\ee 
The first bubble nucleates when $P\sim 1$, which yields a rough
estimate for the nucleation criterion  
$
 S_3(T)/T \sim 4 \ln \left(\frac{M^*_{Pl}}{T}\right) 
 \sim 150 
 $, 
where $T$ is a typical temperature of order the electroweak energy
scale, $T\simeq M_{EW}$. More accurate calculation 
reveals~\cite{Anderson:1991zb}  
\be
S_3(T_c)/T_c \simeq 135.
 \label{NuclCriteria}
\ee

We recall that singlet VEVs $v_S$ and $v_P$ are the  input parameters
of our model.  
   The vacuum $(v,v_S,v_P)$  is the global minimum of the effective potential
$V_{T=0}^{eff}$ in the broken phase  (see discussion in Appendix\,\ref{SectFineTunDim}). At the finite temperature $T\ne
0$, this broken minimum is shifted due to the  termal corrections
(\ref{effpotT1}) 
\be
(v,v_S,v_P) \rightarrow (v_c,S_c, P_c).
\label{BrokenMin}
\ee

In order to find numerically  the profile of the critical bubbles,  
we use the method described in~\cite{Moreno:1998bq,John:1998ip} 
  and later modified in  Ref.\,\cite{Demidov:2006zz}. The procedure
can be  summarized as follows.  
Firstly, for given values of   $v_S$, $v_P$ and $T$ we find
numerically the nearest  minima of the effective potential
$V_{T}^{eff}$ in the symmetric $(0,S_s,P_s)$ and broken $(v_c,S_c,
P_c)$ phases. For technical reason, we shift the
effective potential by a constant to set
$V^{eff}_T(0,S_s,P_s)=0$. Secondly, we construct an anzats for the
bubble wall configurations which interpolate between these two minima
at the temperature $T$.  Next, using this anzats as the first
approximation we numerically find  the absolute minimum of the
functional   
\be 
\mathcal{F}(h,S,P)=4\pi \int_0^{\infty} dr \, r^2 
\left[ E^2_h(r)+E^2_S(r)+E^2_P(r)\right],
\label{FuncEqMot}
\ee 
where $E_h(r)$, $E_S(r)$ and $E_P(r)$ are the
 equations of motion for the bubble wall profiles
$$
E_h(r)\equiv\frac{d^2h}{dr^2}+\frac{2}{r}\frac{dh}{dr}-\frac{\partial V^{eff}_T}{\partial h}=0, 
\qquad E_S(r)\equiv\frac{d^2S}{dr^2}+\frac{2}{r}\frac{dS}{dr} -\frac{\partial V^{eff}_T}{\partial S}=0,
$$
$$
E_P(r)\equiv\frac{d^2P}{dr^2}+\frac{2}{r}\frac{dP}{dr} -\frac{\partial V^{eff}_T}{\partial P}=0.
$$ 
 Note that the critical bubble obeys the following boundary 
conditions
$$
(h(r), S(r), P(r))\Big |_{r=\infty}=(0,S_s,P_s), 
\quad \left(\frac{dh}{dr},\frac{dS}{dr},
\frac{dP}{dr}\right)\Big|_{r=0}=(0,0,0).
$$

\begin{figure}[h]
  \begin{center}
    \includegraphics[width=0.496\textwidth]{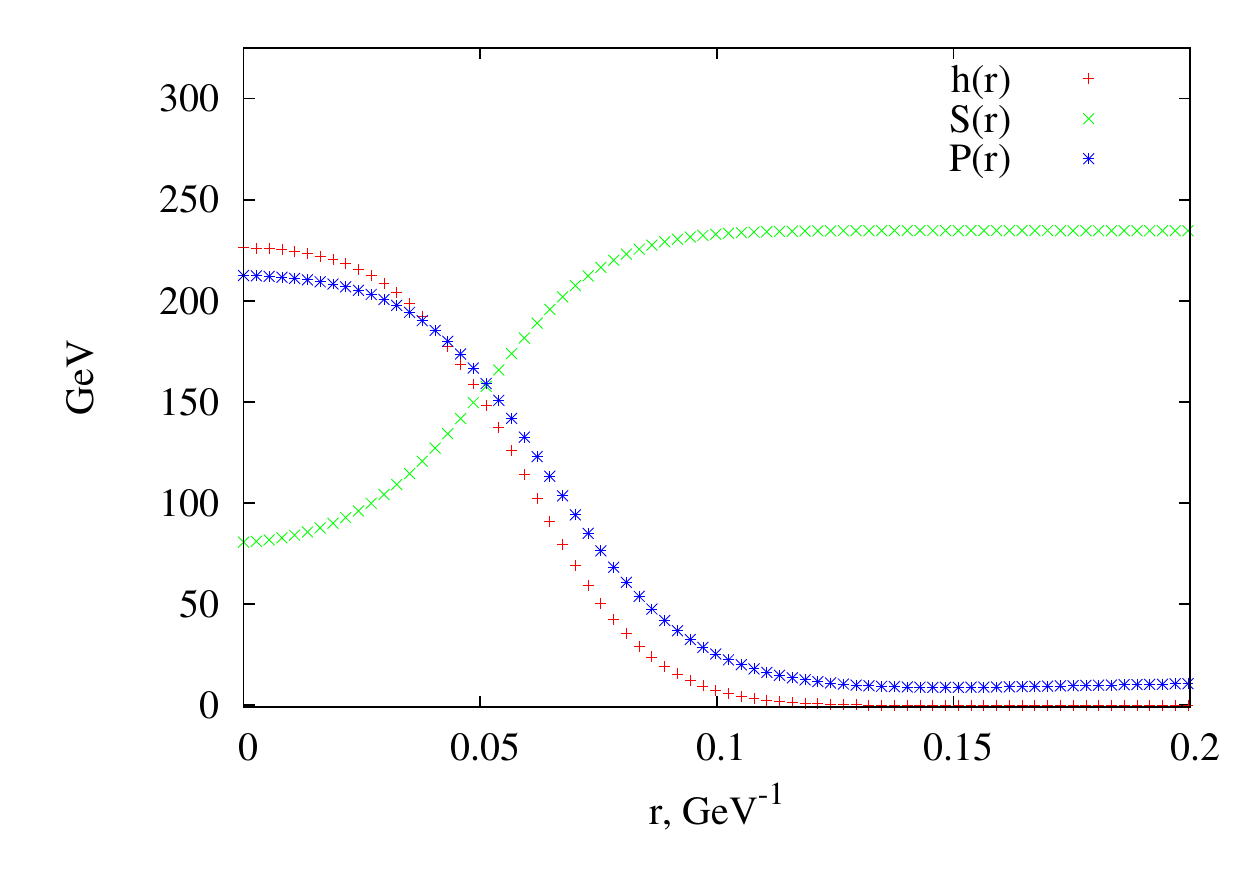}
      \includegraphics[width=0.496\textwidth]{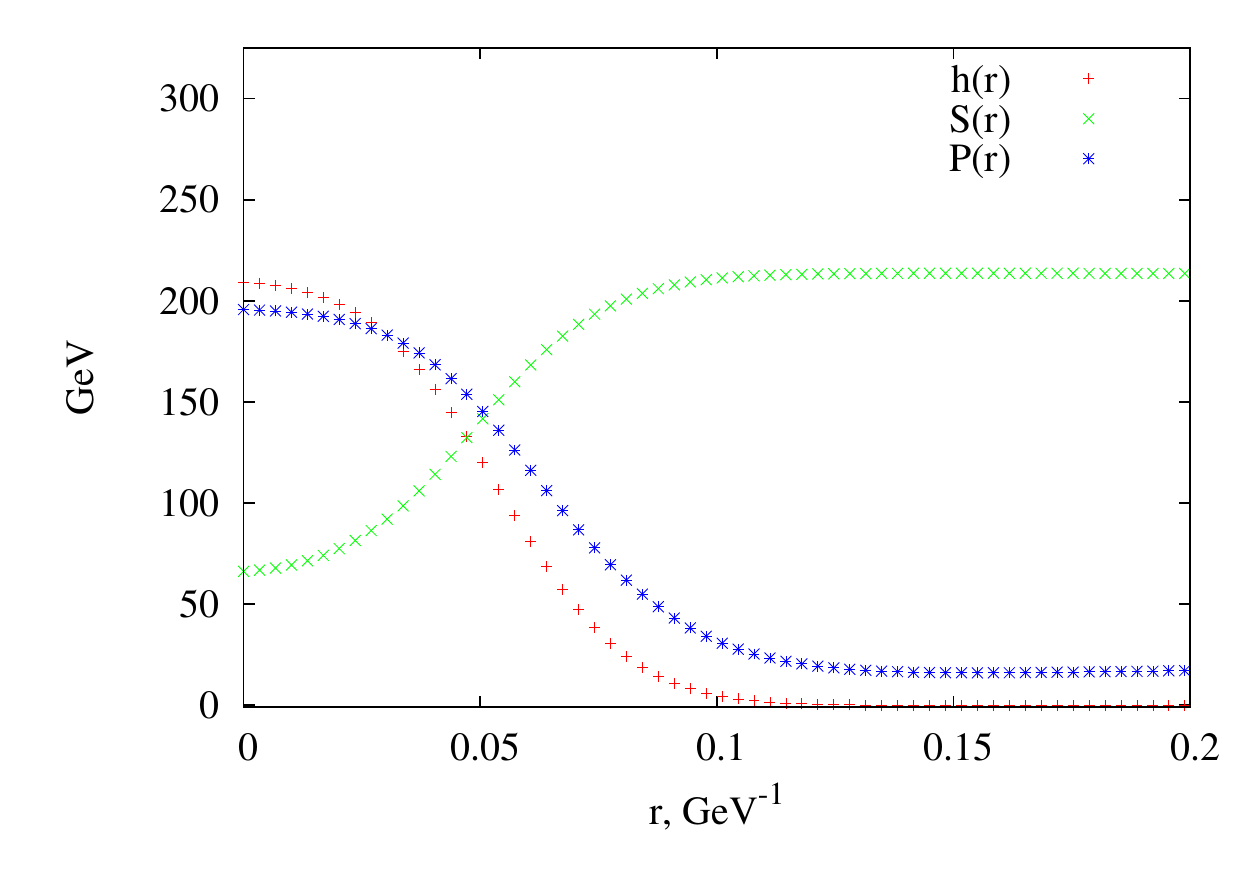}  
    \caption {  The critical bubble profile 
      for the parameter set presented in 
      Tables\,\ref{dimlsscplngs} and 
      \ref{table_values}. Left and right panels 
correspond to Setups $(1)$ and $(2)$, respectively.   
      \label{figure2}}
  \end{center}
\end{figure}

In Fig.\,\ref{figure2} we show dependence of the critical scalar
fields on the radial coordinate for  the selected benchmark models
at their critical temperatures. 
The corresponding values of the relevant physical parameters  are
shown in Table\,\ref{table_values}. All dimensionful parameters in
Table\,\ref{table_values} are in GeV.  
\begin{table}[h]
\begin{center}
\begin{tabular}{|c|c|c|c|c|c|c|c|c|c|c|c|c|}
\hline
 & $v_S$ & $v_P$  & $ {\text Im}\,\mu $  & $T_{c}$ & $v_{c}$ & $S_{c}$ & $P_{c}$ & $S_{s}$ & $P_{s}$ &
$S_{3}/T_c$ 
\\\hline 
Setup\,1 & $60$ & $220$  & $ 151.68  $  &  $67.5$ & $233.29$ & $60.04$ & $219.35$ & $234.53$ & $11.79$ & $136.41$ 
\\\hline
Setup\,2 & $47.5$ & $202.5$  & $ 149.7  $  &  $79.0$ & $220.88$ & $47.59$ & $201.29$ & $213.31$ & $18.65$ & $139.58$ 
\\\hline
\end{tabular}
\caption{\label{table_values} Parameters for the first order EWPT
in the split NMSSM.
}
\end{center}
\end{table}
 We observe considerable change in the values  of the pseudoscalar
  field $P$ in the broken and in the symmetric phases. 
  This will be the source of
$CP$-violation for generation of the baryon asymmetry during the EWPT.

\section{Baryon asymmetry
\label{BaryAsymmSect}}

In this Section we discuss the baryon asymmetry created during the
EWPT in the hot electroweak plasma.  We will closely follow
Ref.~\cite{Huet:1995sh}. To the linear order in chemical  potentials
$\mu_i$, the particle asymmetry  number density for $i$-th
component of plasma reads as  
\be 
 n_i =\frac{1}{6} k_i  \mu_i T^2,
\label{density1}
\ee  
where $k_i$ equals 2 and 1 for massless boson and fermion  degree of 
freedom, respectively. For nonrelativistic particle with mass $m_i$, 
the parameter $k_i$ is suppressed by factor $\exp(-m_i/T)$. 

It is convenient to  introduce the following notations:
$
n_T \equiv n_{t_R}$, $n_Q\equiv  n_{t_L}+n_{b_L}$,  
$n_B \equiv n_{b_R}$,
$
n_{H'}=n_{H^+}+n_{H^0}$,   
$ n_{\tilde{H}_u}=n_{\tilde{H}_u^+}+n_{\tilde{H}_u^0}$, $ 
n_{\tilde{H}_d}=n_{\tilde{H}_d^+}+n_{\tilde{H}_d^0}$,  and similar ones for the
corresponding chemical potentials.  We emphasize that the densities
$n_i$ are local quantities and through the parameters in
\eqref{density1} depend on $z+v_w t$, where $z$ is the  coordinate
perpendicular to the bubble wall, and $v_w$ is the wall velocity.    
The baryon number conservation implies the following relation
$n_B+n_T+n_Q=0$.

In diffusion equations we take into account scattering processes
involving the top Yukawa coupling $y_t \bar{Q} t H$ with the rate
$\Gamma_{Y}$, strong sphaleron transitions with the rate $\Gamma_{ss}
= 6\kappa_{ss}\frac{8}{3}\alpha_{s}^{4}T$, Higgs boson self
interactions with rate $\Gamma_{H}$. We  also include the rate 
$\Gamma_m$ for top quark mass interactions and the rates
$\Gamma_{u,d}$ for Higgs-gaugino-higgsino interactions. In addition,
we  take into account the Higgsino flipping interaction $\tilde{\mu}
\tilde{H}_u \epsilon \tilde{H}_d $ which has the rate $\Gamma_\mu$.  

Following Ref.~\cite{Huet:1995sh}, we write down the set of diffusion
equations in the large $\tan{\beta}$
limit\footnote{ At $\tan{\beta}\sim 1$ the generated baryon
  asymmetry turns out to be suppressed, see Ref.\,\cite{Demidov:2006zz} for
  details.}
\begin{eqnarray}
v_{w}n_{Q}^{\prime} & = & D_{q}n^{\prime\prime}_{Q} - \Gamma_{Y}\left[ 
  \frac{n_{Q}}{k_{Q}} - \frac{n_{T}}{k_{T}} - \frac{n_{H} +
  n_{h}}{k_{H_{1}}}\right] - \Gamma_{m}\left[\frac{n_{Q}}{k_{Q}} -
  \frac{n_{T}}{k_{T}}\right]\label{first} \\
&& - 6\Gamma_{ss}\left[2\frac{n_{Q}}{k_{Q}}
  - \frac{n_{T}}{k_{T}} + 9\frac{n_{Q} + n_{T}}{k_{B}}\right],
 \nonumber  \\
v_{w}n_{T}^{\prime} & = & D_{q}n_{T}^{\prime\prime} + \Gamma_{Y}\left[ 
  \frac{n_{Q}}{k_{Q}} - \frac{n_{T}}{k_{T}} - \frac{n_{H} +
  n_{h}}{k_{H_{1}}}\right] + \Gamma_{m}\left[\frac{n_{Q}}{k_{Q}} -
  \frac{n_{T}}{k_{T}}\right]\label{second} \\
&& + 3\Gamma_{ss}\left[2\frac{n_{Q}}{k_{Q}}
  - \frac{n_{T}}{k_{T}} + 9\frac{n_{Q} + n_{T}}{k_{B}}\right], \nonumber\\
v_{w}n_{H}^{\prime} & = & D_{h}n_{H}^{\prime\prime} + \Gamma_{Y}\left[ 
  \frac{n_{Q}}{k_{Q}} - \frac{n_{T}}{k_{T}} - \frac{n_{H} +
  n_{h}}{k_{H_{1}}}\right] - \Gamma_{H}\frac{n_{H} + n_{h}}{k_{H_{1}}} 
  +  S_{u} - S_{d}, 
  \label{third} \\ 
v_{w}n_{h}^{\prime} & = & D_{h}n_{h}^{\prime\prime} + \Gamma_{Y}\left[
  \frac{n_{Q}}{k_{Q}} - \frac{n_{T}}{k_{T}} - \frac{n_{H} +
  n_{h}}{k_{H_{1}}}\right] - 2\l[\frac{n_{H}}{k_{H}} +
  \frac{n_{h}}{k_{h}}\r]\Gamma_{\tilde{\mu}}\label{fourth} \\
  && - \frac{n_{H} + n_{h}}{k_{H_{1}}}\Gamma_{H}  + S_{u} +
  S_{d}, \nonumber
\end{eqnarray}
 which are written for combinations of the Higgs bosons and higgsino
densities $n_{h} = n_{H^{\prime}} + n_{\tilde{H}_{u}} + 
n_{\tilde{H}_{d}}$ and $ n_{H} = n_{H^{\prime}} + 
n_{\tilde{H}_{u}} - n_{\tilde{H}_{d}}$.
Here  the prime and double prime denote the first and second
derivatives with respect to variable $z$. 
In equations\,\eqref{first}-\eqref{fourth}, we set   
 $\Gamma_{\tilde{\mu}} \equiv \Gamma_{\mu} + \Gamma_{d}$,  
statistical factors  are defined by
$$
k_{H_{1}} = 2(k_{H^{\prime}} + k_{\tilde{H}_{u}}),\;\;\;
k_{H} = \frac{2(k_{H^{\prime}} +
  k_{\tilde{H}_{u}})k_{\tilde{H}_{d}}}{k_{\tilde{H}_{d}} -
  k_{H^{\prime}} -  k_{\tilde{H}_{u}}},\;\;\;
k_{h} = \frac{2(k_{H^{\prime}} +
  k_{\tilde{H}_{u}})k_{\tilde{H}_{d}}}{k_{\tilde{H}_{d}} +
  k_{H^{\prime}} +  k_{\tilde{H}_{u}}},
$$
and 
$CP$ -violating sources $S_u$ and $S_d$ are discussed below in due course. 
 We assume that the diffusion coefficients $D_q$
are the same for all quarks, and $D_h$ are the same for all Higgs bosons
and higgsinos. Using the approach advocated in 
Ref.~\cite{Carena:1997ki}, we eliminate $n_{T}$ and $n_Q$ from
Eqs.\,(\ref{first})-(\ref{fourth}) by substituting the relations  
\be
\label{QT}
n_{T} = - \frac{k_{T}\l(2k_{B} + 9k_{Q}\r)}{k_{H_{1}}\l(9k_{Q} +
  9k_{T} + k_{B}\r)}\l(n_{H} + n_{h}\r),\;\;
n_{Q} = \frac{k_{Q}\l(9k_{T} - k_{B}\r)}{k_{H_{1}}\l(9k_{Q} +
  9k_{T} + k_{B}\r)}\l(n_{H} + n_{h}\r),
\ee
which follow from the assumption  that both the top Yukawa
interactions and the strong sphalerons are in equilibrium.
The  resulting equations for  $n_{H}$ and $n_{h}$ are collected
in Ref.~\cite{Demidov:2006zz}. It follows from (\ref{QT}),  that the 
left-handed fermion  density can be recasted in the following form
\begin{multline}
\label{n_L}
n_{Left} = n_{Q_1} + n_{Q_2} + n_{Q_3} = 4n_{T} + 5n_{Q} \\ =
 \frac{5k_{Q}k_{B} +   8k_{T}k_{B} - 9k_{Q}k_{T}}{k_{H_{1}}\l(k_{B}
 +  9k_{Q} + 9k_{T}\r)}\l(n_{h} +   n_{H}\r) \equiv A_t\cdot
 (n_{h} + n_{H}). 
\end{multline}
The statistical factors are 
$k_{Q} = 6$, $k_{T} = 3$, $k_{B} =
3$,  $ k_{H^{\prime}} = 4$, 
$k_{\tilde{H}_{u}} = k_{\tilde{H}_{d}} = 2$ and hence the constant
$A_t$  is equal to zero~\cite{Giudice:1993bb}.   
It was shown in Ref.~\cite{Carena:2004ha}, that one-loop
corrections to statistical  coefficients $k_i$ give non-zero
value of $A_t$, namely 
\be
\label{n_L1}
A_t = \frac{3}{2k_{H_{1}}}\l(-\frac{3y_{t}^{2}}{8\pi^{2}}\r).
\ee
The  baryon asymmetry  obeys the following equation
\cite{Carena:1997ki}  
\be
v_{w}n_{B}^{\prime}(z) = -\theta(-z)\left[n_{F}\Gamma_{ws}n_{Left}(z) +
  {\cal R}n_{B}\r], 
  \label{BaryonEq1}
\ee
where $\Gamma_{ws} = 6\kappa_{ws}\alpha_{w}^{5}T$ is  the weak
sphaleron rate with $\kappa_{ws} = 20\pm 2$ \cite{Moore:1999fs}. The   
relaxation coefficient ${\cal R}$ is given by \cite{Cline:1997vk}  
${\cal R} = \frac{5}{4}n_{F}\Gamma_{ws}$,
and $n_{F}$ is the number of  generations, 
$n_{F} = 3$. Here the domain $z<0$ corresponds to the symmetric
  phase. The solution to Eq.~(\ref{BaryonEq1}) reads
\be
\label{n_L2}
n_{B} =-\frac{n_{F}\Gamma_{ws}}{v_{w}}\int_{-\infty}^{0}
dzn_{Left}(z)e^{z{\cal R}/v_{w}}.
\ee

In the split NMSSM,  $CP$-symmetry gets violated spontaneously 
while the bubble walls expand in the hot plasma.  
Indeed,    the  main 
source of $CP$-violation is associated with the  
complex chargino  mass matrix
\be
\label{charg_matr}
M_{ch} = 
\l(
\begin{array}{cc}
M_{2} & \frac{1}{\sqrt{2}}\tilde{g}_{u}h(z) \\
\frac{1}{\sqrt{2}}\tilde{g}_{d}h(z) & \tilde{\mu}(z)
\end{array}
\r),
\ee
where we define the spatially-dependent effective higgsino
mass parameter as follows 
\be 
\tilde{\mu}(z) = \mu + 
\kappa\l(S(z) + iP(z)\r)/\sqrt{2}.
\label{CharginoMassParam}
\ee 
In the above expressions,  $h(z)$, $S(z)$ and 
$P(z)$ are the kink approximations of the bubble walls
 \cite{Carena:1997ki}
\begin{gather}
h(z) = \frac{1}{2}v_{c}\l(1 - \tanh{\left[\alpha\l(1 -
    \frac{2z}{L_{w}}\r)\right]}\r),\\
\l(
\begin{array}{c}
S(z) \\
P(z) \\
\end{array}
\r)
= 
\l(
\begin{array}{c}
S_{c} \\
P_{c} \\
\end{array}
\r)
-
\frac{1}{2}\l(
\begin{array}{c}
\Delta S \\
\Delta P \\
\end{array}
\r)
\l(1 + \tanh{\left[\alpha\l(1 -
    \frac{2z}{L_{w}}\r)\right]}\r),
\end{gather}
here $v_c$, $S_c$ and $P_c$ are the critical values of the scalar
fields (see, e.g. Table\,\ref{table_values}), $\Delta S\equiv S_c-S_s$ and
$\Delta P \equiv P_c-P_s$.  We set velocity of the bubble wall equal
to $v_w=0.1$, the coefficient $\alpha$ is taken to be $3/2$.  The
bubble wall width 
$L_w$ may be chosen in the range $5/T_c <L_w < 30/T_c$ consistent with
the special study~\cite{Huber:2000mg} and the WKB 
thick-wall restriction, $L_w T_c > 1$.
 
Following Ref.~\cite{Huber:2000mg}, we define the rates 
in Eqs.~(\ref{first}-\ref{fourth}) as 
$$
\Gamma_{H} = 0.0036T\theta\l(z - 0.5L_{w}\r),\quad
\Gamma_{m} = 0.05T\theta\l(z - 0.5L_{w}\r),\quad
\Gamma_{\tilde{\mu}}=0.1 T,
$$
with $\theta$ being the step function,  
 and choose the diffusion coefficients in the  
 form  $D_q = 6/T$ and $D_h = 110 /T$.

 \begin{figure}[t]
\begin{center}
\includegraphics[width=0.8\textwidth]{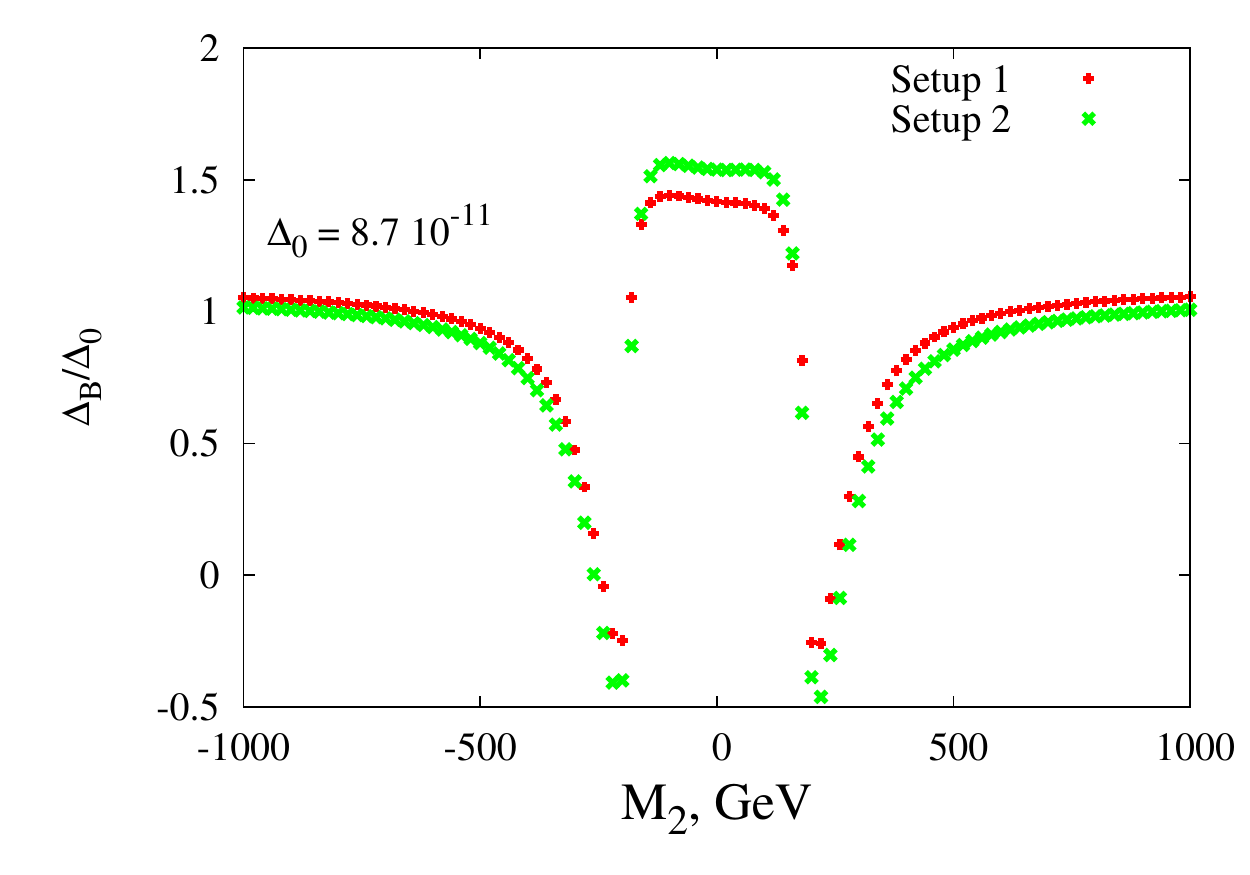}
\caption { Plot of $\Delta_B/\Delta_0$ versus gaugino mass 
parameter $M_2$ for the 
parameter sets presented in Tables\,\ref{table_values} and
\ref{dimlsscplngs}.  \label{figure3}}
\end{center}
\end{figure}   

We use the expressions for $CP$-violating sources  $S_d$ and
$S_u$  from Ref.\,\cite{Huber:2000mg} and numerically solve the
  set of diffusion equations for $n_h(z)$ and $n_H(z)$. 
   Then, we
  calculate the asymmetry of left fermions using Eq.~(\ref{n_L}) and
  by evaluating the integral~(\ref{n_L2}) we obtain the baryon
  asymmetry generated during EWPT.

Let us consider the baryon-to-entropy ratio $\Delta_B=n_B/s$ with
the entropy density 
$$s=2\pi^2g_{eff}T^3/45,$$ 
where $g_{eff}$ is the effective number 
of relativistic degrees of freedom at $T_c$. 
In Fig.\,\ref{figure3} 
we show  dependence of the baryon asymmetry 
$\Delta_B/\Delta_0$ on gaugino mass $M_2$ 
for different 
values of the wall thickness: namely, we take 
 $L_w=7/T_c$ and $L_w=5/T_c$ for  Setup 1 and 2, 
respectively.  
The  value
$\Delta_0=8.3 \times 10^{-11}$  corresponds to 
$n_B/n_\gamma =6.2 \times 10^{-10}$ consistent with present
measurements \eqref{BaryonAsymm1}.

It follows from Fig.~\ref{figure3} that baryon asymmetry 
$\Delta_B$ is of order $ \Delta_0$  for large  $M_2 \gsim 1$ TeV.
In this case, the heaviest chargino $\chi^{+}_2$
(wino-like) decouples from the plasma, $|m_{\chi^+_{2}}|\simeq M_2$,
and the lightest chargino (higgsino-like) acquires the mass
$|m_{\chi^+_{1}}|$, which is determined by the effective
$\tilde{\mu}(z)$-parameter in (\ref{CharginoMassParam}). Thus, the baryon
asymmetry is generated due to the spontaneous  $CP$-violation in the
broken  (and symmetric) phase. Detailed calculation of  $CP$-violating  sources  
\cite{Huber:2000mg} reveals that $S_u$ and $S_d$ gain contributions
which are proportional to the second derivative of ${\text Im}\,\tilde{\mu}(z)$ with respect to $z$ coordinate. This means that
baryon asymmetry $\Delta_B/\Delta_0 $ is rather sensitive to the  
effective parameter, $${\text Im}(\tilde{\mu}'') \sim  \kappa \Delta
P/L^2_w.$$   In our numerical analysis, we tune the wall thickness
$L_w$ to obtain $\Delta_B/\Delta_0 \sim 1$ as $M_2 \rightarrow \infty$. 
At the same time from the very beginning we choose sufficiently large
coupling $\kappa$ (by taking large $\lambda$) and pseudoscalar
VEV gradient $\Delta P = P_c-P_s$ 
and large value of $\tan\beta$. These features select models which are 
interesting for the realistic electroweak baryogenesis. As we will see  in
Section\,\ref{EDMLHC} the latter condition is also preferred by 
present electron's EDM  constraints. From Fig.~\ref{figure1} we see
that large values of $\lambda$ and $\tan\beta$ require moderate value
of the splitting scale $M_S$, which hardly can be larger than
$12-15$~TeV. 

In our analysis we can evaluate the baryon asymmetry in the limit $n_h
\gg n_H$, following  the approach,  presented in
Ref.~\cite{Carena:2000id}. In this approximation the set of diffusion
equations  (\ref{first}-\ref{fourth}) reduces to a single
equation on $n_h$, and baryon asymmetry ratio, $\Delta_B/\Delta_0$,
can be 
estimated analytically. In the limit when heaviest chargino decoupled,
$m_{\chi_{2}^+} \approx M_2 \approx$ 1 TeV, one finds
\be
\frac{\Delta_B}{\Delta_0} \approx 5.5 \cdot 10^2 
\left(\frac{m_{\chi_1^+}}{T_c} \right)^2 \exp\left(-\frac{m_{\chi_1^+}}{T_c} \right) \frac{1}{(L_w T_c)^2}.
\label{BaryonRationAnalyt}
\ee
For $L_w = 5/T_c$, $T_c = 80$ GeV and $m_{\chi_1^+} =239$ GeV 
this yields $\Delta_B/\Delta_0  \approx 10$. An order-of-magnitude  
discrepancy between the numerical, 
$\Delta_B/\Delta_0\approx 1$, and analytic results  
(\ref{BaryonRationAnalyt}),  is due to the 
approximations which have been made for solving equation for 
$n_h$ in the analytically approach.  
 Let us note that here we estimate baryon asymmetry originated from chargino
  sector only. $CP$-violating sources from neutralino sector can change
  the calculated value of the asymmetry by a factor of order one.

\section{EDM constraints and light chargino phenomenology
\label{EDMLHC}}
 
In this Section we address  some phenomenological implementation
of the results discussed above. To begin with, we emphasize that
current constraints on electric  dipole moments of the electron and
neutron provide strong limits for $CP$-violating physics in the split
NMSSM. There are three relevant contributions to the EDM of electron
or light quark 
\cite{Arkani-Hamed:2004yi,Giudice:2005rz},
$$
d_f = d_f^{H\gamma}+d_f^{HZ}+d_f^{WW}, 
$$
where $d_f^{H\gamma}$, $d_f^{HZ}$ and $d_f^{WW}$ are the partial EDMs
of fermion (lepton or quark), related to the exchange of $H\gamma$,
$HZ$ and $W^+W^-$ bosons, respectively. General expressions for the
electron's EDM $d_e$ and neutron's EDM $d_n$  were derived   in
Ref.~\cite{Giudice:2005rz}. The values of $d_e$ and $ d_n$ depend on
chargino, $m_{\chi^+_{i}} \, (i=1,2)$,  and neutralino,
$m_{\chi^0_{j}}\, (j=1,5)$,   masses as well as their mixing
matrices\footnote{We recall that  neutralino state $\chi^0_j$ in
split NMSSM is  determined by the  mixing of neutral bino
$\tilde{B}^0$, wino $\tilde{W}^0$, higgsino $\tilde{H}^0_{u,d}$ and  
sing lino $\tilde{n}$ states.}. 
 
\begin{figure}[!htb]
\includegraphics[width=0.5\textwidth]{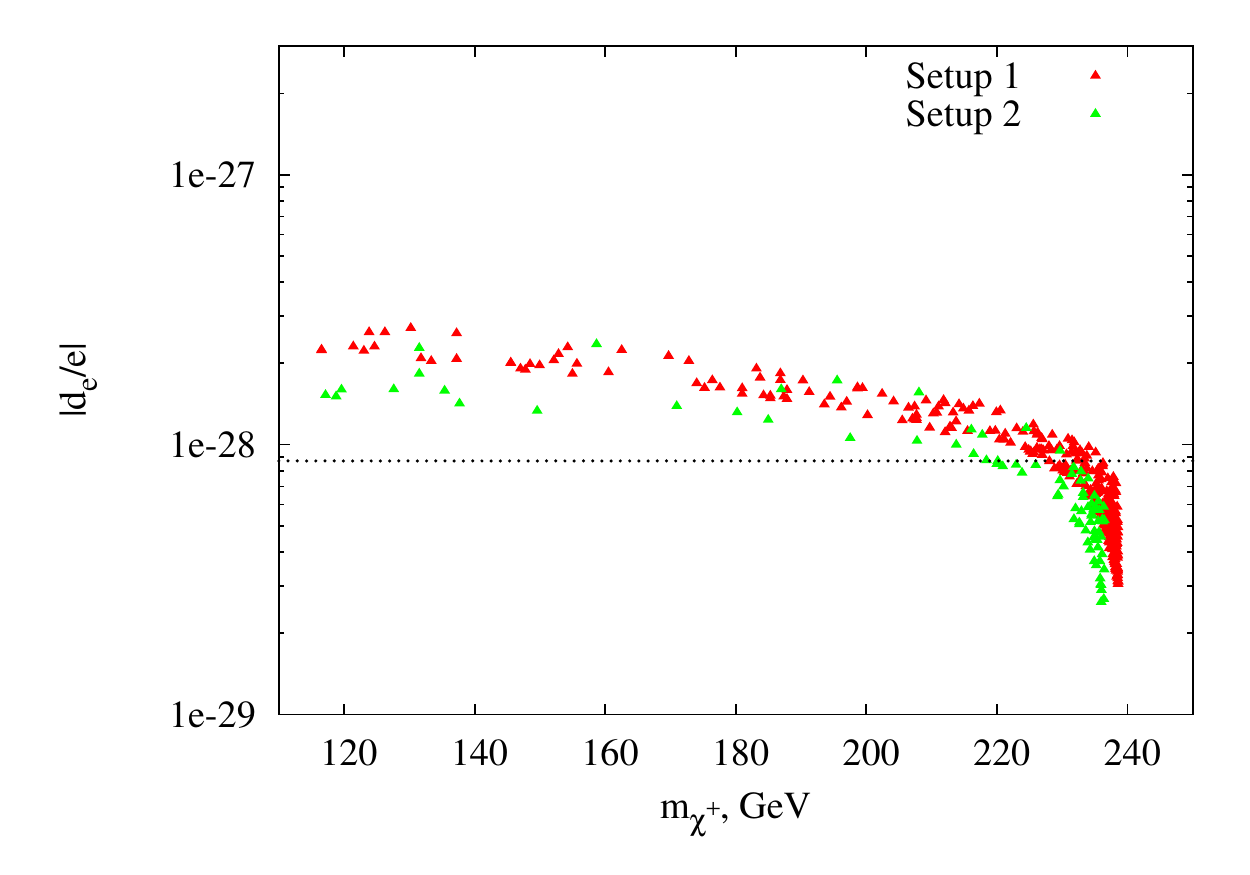}
\includegraphics[width=0.5\textwidth]{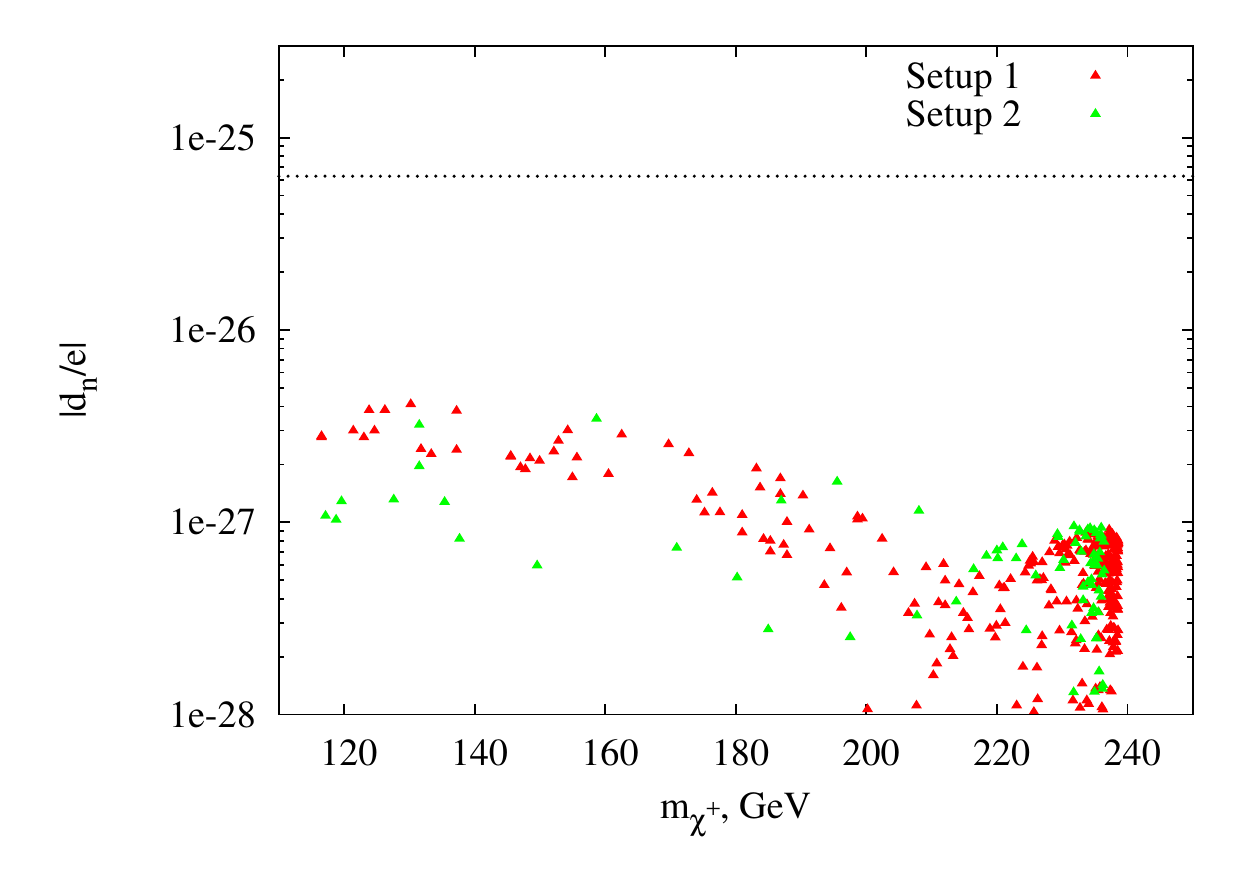}
\caption {Left panel: the EDM of electron versus the lightest
  chargino  
mass $m_{\chi^+_1}$. Dotted lines represent the current experimental 
bound  $|d_e/e| < 8.7 \times 10^{-29}$\,cm. Right panel: the neutron's 
EDM  with upper limit $|d_n/e|<3.0 \times 10^{-26}$\,cm. The relevant
couplings, $\mu$-terms and both singlet VEVs $v_S$ and $v_P$ at $T=0$
are given  in Tables\,\ref{dimlsscplngs} and \ref{table_values}.  
 \label{figure4}}
\end{figure}  
 
The most stringent upper limit on EDM of the electron, $|d_e/e| < 8.7
\times 10^{-29}$\,cm at 90$\%$ CL,  was obtained by ACME collaboration
\cite{Baron:2013eja}. The current bound on neutron's EDM
is  $|d_n/e|<3.0 \times 10^{-26}$ cm at 90 $\%$ CL
\cite{Baker:2006ts}. In order to perform the numerical analysis for
EDMs, we randomly scan over the following parameter
space $0<M_1,M_2<1000$~GeV.
In Fig.\,\ref{figure4} we show  dependence of $|d_e/e|$ on the lightest
chargino mass $m_{\chi^+_1}$. One can see 
from the left panel of Fig.~\ref{figure4}
that chargino masses  in the ranges  $225\, \mbox{GeV}
<m_{\chi_1^+}<239$~GeV and $220\, \mbox{GeV}
<m_{\chi_1^+}<235$~GeV are allowed for the Setup 1 and 2,
respectively.  We check that all these points correspond to large
  (about 1~TeV) values of $M_2$ and hence allow for correct value of
  the BAU. 
The numerical results for neutron EDM are shown on right panel of 
Fig.\,\ref{figure4}.  One can see from Fig.\,\ref{figure4}  
that predictions for the neutron EDMs satisfy the current experimental
bound  in all selected models.  
 For the Setup 1 we present  an  examples of chargino and neutralino  
mass   spectra   which are consistent with the EDM bounds
in    Fig.~\ref{figure4} for  $M_1=300$~GeV and $M_2=1$~TeV
\begin{itemize}
 \item $m_{\chi_1^+}=238.4$ GeV, $m_{\chi_2^+}=1006.8$ GeV,
 \item  $m_{\chi_1^0}=133.9$ GeV, $m_{\chi_2^0}=220.5$ GeV,  
 $m_{\chi_3^0}=268.0$ GeV,
  $m_{\chi_4^0}=341.9$ GeV,   $m_{\chi_5^0}=1006.9$ GeV.
\end{itemize}
We find in this case, that LSP is singlino-like state  with the
  mass $m_{\chi_1^0}=133.9$ GeV.  The dominant decay channel of
  the lightest chargino is  $\chi_1^+\rightarrow \chi_1^0 W^+$, 
  which can be used to test split NMSSM model. In our analysis, we
  checked  that the models satisfying EDM bounds  are in
  agreement with the present  
CMS\,\cite{Khachatryan:2014qwa} and ATLAS\,\cite{Aad:2014vma} 
limits on chargino-neutralino production at LHC without light
sleptons. Therefore, the split NMSSM is a phenomenologically viable
and cosmologically attractive model which can be probed at the LHC run
with $pp$ collision energy of 13\,TeV (and 14\,TeV).

\section{Conclusion}

In this paper we revisit scenario of non-minimal split supersymmetry
with possibility of realistic electroweak baryogenesis. It is a
realization of split supersymmetry in the framework of NMSSM and
contains at the electroweak scale, apart from minimal split
supersymmetry particle content, singlet scalar and pseudoscalar
states. We observed that within the phenomenologically allowed domain
of the parameter space with the mass of the Higgs boson equal to
125~GeV it is possible to find particular models in which the strongly first order 
electroweak phase transition can be realized and moreover the needed
amount of the baryon asymmetry of the Universe is generated.
 These models predict existence of light chargino state required
  for successful baryogenesis. We also find relatively light LSP with
  large  admixture of singlino like state. Therefore, it can be
  considered as a potential dark matter candidate as suggested in
  Ref.\,\cite{Demidov:2006zz}. Predictions for the
  electric dipole moment of electron in these models are found  to be
 about or somewhat larger 
than $2-3\cdot 10^{-29}e$~cm which is only by factor 3-4 smaller
  than the current upper limit on this quantity. This makes the
  searches for EDMs a promising tool to probe the split NMSSM.

The work was supported by the RSF grant 14-22-00161.
 



\appendix

\section{One loop corrections to Higss mass in split NMSSM
\label{Appendix1LoopHiggsMass}} 

In this appendix we  calculate one-loop RG corrections to the mass
  of Higgs boson in  split NMSSM scenario following
  Refs.~\cite{Binger:2004nn,Degrassi:2009yq}. In particular, in
  Ref.~\cite{Degrassi:2009yq}  
the radiative corrections to the Higgs mass were calculated in the
NMSSM in Ref.~\cite{Degrassi:2009yq}, while they were derived
explicitly in split MSSM in Ref.~\cite{Binger:2004nn}. However,  
split MSSM  computations~\cite{Binger:2004nn} can be straightforwardly
extended to the  split NMSSM case by taking  into account the
radiative corrections from scalars, charginos and neutralinos,   
\be 
(m^{pole}_{h})^2=(m^{tree}_{h})^2(\mub) +\delta^{SM}_h(\mub) + \delta^{(S,P)}_{h}(\mub) + \delta^{(C,N)}_{h}(\mub),
\label{OneLoopHiggsMass1}
\ee
where $(m_h^{tree})^2=\tilde{\lambda}(\mub) v^2$ is the three level
Higgs boson mass at $\mub$ scale (dimensional  renormalization scale
in $\overline{\mbox{MS}}$ scheme); the remnant one-loop corrections  
in (\ref{OneLoopHiggsMass1}) are defined below in
Sections~\ref{ScalarsOneLoop} and~\ref{GauginoHiggsinoOneLoop}. 
  We use the experimental value of the Higgs pole mass
(\ref{OneLoopHiggsMass1}) to plot the figures for the allowed region
of split NMSSM parameters in the main text. 

\subsection{Tree level potential of scalar sector in
 the broken phase \label{ScalarsOneLoop}}

Applying the general results of Ref.~\cite{Degrassi:2009yq} we
rewrite~(\ref{gener_poten}) in the broken phase,
\be \
 H=(\phi_1+v)/\sqrt{2}, 
\qquad   N=(\phi_2+v_S+i(\phi_3+v_P))/\sqrt{2}, 
\label{BrPhScFields}
\ee
 where we denote perturbations of the scalar fields about the vacuum 
as $(\phi_1, \phi_2, \phi_3) = (h, S, P)$.
Then, substituting (\ref{BrPhScFields}) 
into (\ref{gener_poten}) and using minimization conditions
(\ref{msquared1}-\ref{msquared2}) at the tree level, 
 one can obtain  
\be 
\mathcal{L}_{V}  \supset - 
\sum_{ijkl} \lambda_{\phi_i \phi_j \phi_k \phi_l} 
\phi_i \phi_j \phi_k \phi_l - 
\sum_{ijk} \lambda_{\phi_i \phi_j \phi_k} \phi_i \phi_j \phi_k    
- \sum_{ij}  \frac{1}{2} m^2_{\phi_i \phi_j} \phi_i \phi_j.
    \label{VScTreeBrPh}
\ee
The quartic and trilinear couplings which are relevant for the
calculation of  the Higgs boson self energy and tadpoles in the
scalar sector of  the split NMSSM can be written as  
\be 
\lambda_{\phi_1\phi_1\phi_1\phi_1}=\frac{1}{8}\tilde{\lambda}, \qquad
\lambda_{\phi_1\phi_1\phi_2\phi_2} =\frac{1}{12}(\kappa_1+\kappa_2), 
\qquad \lambda_{\phi_1\phi_1\phi_3\phi_3} =
\frac{1}{12}(\kappa_1-\kappa_2), 
\ee
\be
\lambda_{\phi_1\phi_1\phi_1}=\frac{1}{2}\tilde{\lambda} v, \qquad \lambda_{\phi_1 \phi_2 \phi_2}= \frac{1}{3} (\kappa_1+\kappa_2)v, \qquad
 \lambda_{\phi_1 \phi_3 \phi_3}= \frac{1}{3} (\kappa_1-\kappa_2)v,
\ee
\be
\lambda_{\phi_1\phi_3\phi_2}=\lambda_{\phi_1\phi_2\phi_3}=0. 
\ee
The parameters of  the scalar squared mass matrix read
\be 
m_{\phi_3\phi_3}^2=(\kappa_1-\kappa_2)v^2+\lambda_N(3 v_P^2+v_S^2)
-\tilde{\lambda} (v_P^2+v_S^2), 
\ee
\be
\quad m^2_{\phi_2\phi_3}=m_{\phi_3\phi_2}^2= -\sqrt{2} 
\tilde{A}_{k} v_P +2 \lambda_N v_P v_S.
\ee 
\be 
m^2_{\phi_1\phi_1}=\tilde{\lambda}v^2, \qquad m_{\phi_2\phi_2}^2=(\kappa_1+\kappa_2)v^2+\lambda_N(v_P^2+3v_S^2)+
\left(-\tilde{\lambda}+\tilde{A}_k/(\sqrt{2}v_S)\right)
 (v_P^2+v_S^2),
\ee 
\be 
 m^2_{\phi_1 \phi_3}=m^2_{\phi_3 \phi_1}=m^2_{\phi_1 \phi_2}=m^2_{\phi_2 \phi_1}=0.
 \label{OffdiagMass}
\ee 
One should diagonalize  its $2 \times 2$ submatrix for the  
singlets $m^2_{\phi_i \phi_j}$, with $i,j=2,3$, since  off-diagonal
mixings of $\phi_2$ and $\phi_3$ with the Higgs field $\phi_1$ are set
to be zero (\ref{OffdiagMass}) 
 (see  also discussion before Eq.~(\ref{A1A2})). We denote  
  the  singlet eigenstates by $h_i$ and diagonalize
 $m^2_{\phi_i\phi_j}$  by an orthogonal matrix $R_{ij}$, such that  
 \be 
 h_i = R_{ij} \phi_j.
 \ee
 The couplings that enter the calculation of the Higgs boson
 mass radiative  corrections can be expressed as   
 \be 
 \lambda_{\phi_i\phi_j h_k h_l}= 6\, R_{k a}\, R_{lb} \,
 \lambda_{\phi_i \phi_j \phi_a \phi_b}, \qquad 
 \lambda_{\phi_i h_k h_l} = 3\,  R_{k a} \, R_{lb} \,
 \lambda_{\phi_i \phi_a \phi_b}.
 \ee 
Following the prescription of Ref.~\cite{Degrassi:2009yq}  we write 
down one-loop contribution of the scalar singlets to the Higgs boson
mass\,
\footnote{Here only scalars $\phi_2$ and $\phi_3$ are taken into account;  
all signs and prefactors correspond to notations from
Ref.~\cite{Degrassi:2009yq}.} 
\be 
\delta^{(S,P)}_h(m_h,\overline{\mu}) =  \frac{1}{v}\, T_h^{(S,P)}(\overline{\mu})-
\Pi_h^{(S,P)} (m_h,\overline{\mu}),
\label{HiggsSPCorr}
\ee
where the Higgs boson self energy is
\be
16 \pi^2 \,  \Pi^{(S,P)}_h(p^2,\overline{\mu})=
\sum_{k=2,3} 2 \lambda_{\phi_1 \phi_1 h_k h_k} A_0(m_{h_k})
+ \sum_{k,l=2,3}  2 \lambda_{\phi_1 h_k h_l} \lambda_{\phi_1 h_k h_l}
B_0(p,m_{h_k}, m_{h_l}), 
\ee
and  the tadpole contributions are 
\be
16 \pi^2 T^{(S,P)}_h(\mub)= \sum_{k=2,3} \lambda_{\phi_1 h_k h_k} 
A_0(m_{h_k}). 
\ee
The loop functions $A_{0}(m)$ and $B_{0}(p, m_1,m_2)$ depend 
on the renormalization scale $\mub$ and can be written in the form
\be 
A_0(m)= m^2\l(C_{UV}+1 - \ln \frac{m^2}{\mub^2} \r), \quad
B_{0}(p,m_1,m_2) = C_{UV} - \ln \frac{p^2}{\mub^2} 
-f_B(x_+)-f_B(x_-),
\ee
where $C_{UV}=1/\epsilon-\gamma_E+\ln 4\pi $, 
$ f_B(x)=\ln (1-x)-x \ln(1-x^{-1})-1$  with
\be
 \qquad x_{\pm} =
 \frac{s\pm\sqrt{s^2-4p^2(m_1^2-i \epsilon)}}{2p^2}, 
 \quad  s=p^2-m_2^2+m_1^2.
\ee
A simplified formula for $B_{0}(p^2,m_1,m_2)$ at $p^2=0$ read \cite{Pierce:1996zz},
\be
B_{0}(0,m_1,m_2)=-\ln \frac{M^2}{\mub^2} +1 +\frac{m^2}{m^2-M^2} 
\ln \frac{M^2}{m^2},
\label{B0simpl} 
\ee 
where $M=\mbox{max}(m_1,m_2)$ and $m=\mbox{min}(m_1,m_2)$.
\subsection{Chargino-neutralino sector of split NMSSM
\label{GauginoHiggsinoOneLoop}}
The Lagrangian of interest for chargino/neutralino sector is  
\be 
-\mathcal{L}^{(C, N)}_{int}=-\frac{1}{2}h \,  \bar{\chi}_i^0 
\l(R_{(ij)}^{N*}P_L + R_{(ij)}^{N}P_R \r) \chi_j^0
\label{ChargNeutInter}+ 
\ee 
$$
+ \l(g \bar{\chi}^+_i \gamma^\mu \l(C_{ij}^R P_R +C_{ij}^L P_L\r)
\chi^0_j W_\mu^+  + 
\frac{1}{\sqrt{2}} \bar{\chi}^+_i \l(R_{ij}^C P_R + L_{ij}^C P_L
 \r) \chi^+_j h  + h.c. \r),
$$
where 
\be 
R_{ij}^N=(\tilde{g}_u N_{i2} -\tilde{g}_u'N_{i1})N_{j4}-
(\tilde{g}_d N_{i2} -\tilde{g}_d'N_{i1})N_{j3}+\sqrt{2}(\lambda_u N_{i4}-\lambda_d N_{i3})N_{j5}
\label{RijN}
\ee 
\be
R_{(ij)}^N=\frac{1}{2}(R_{ij}^N+R_{ji}^N), \qquad 
R_{ij}^C = (L_{ji}^C)^* =  \tilde{g}_u^* V_{i2} U_{j1} + \tilde{g}_d^* V_{i1} 
U_{j2}, 
\ee
\be
 C_{ij}^L=N_{i2}V_{j1}^*-\frac{1}{\sqrt{2}}N_{i4}V_{j2}^*, \qquad  
 C_{ij}^R=N_{i2}^*U_{j1}+\frac{1}{\sqrt{2}}N_{i3}^*U_{j2}. 
\ee
Following Ref.~\cite{Binger:2004nn}, let us consider the contribution
of chargino and neutralino to  the Higgs boson mass at one-loop
level, 
\be 
\delta^{(C,N)}_{h}= \Sigma_{h}^{(C,N)}(m_h,\mub)+
\frac{1}{v} T^{(C,N)}_h(\mub) +
 \frac{\tilde{\lambda} v^2}{m_W^2} \Pi^{(C,N)}_{WW}(0,\mub)
 \label{ChNeutCorr}
\ee
where $T_{h}^{(C,N)}=T^{(C)}_h+T^{(N)}_h$ is the Higgs boson tadpole
contribution  which involves terms   
\be
 16 \pi^2\, T^{(C)}(\mub) = -2\sqrt{2}\sum_{i=1}^2 \mbox{Re}
\l[R_{ii}^C M^C_i A_{0}(M^C_i) \r], 
\ee
\be 
 16 \pi^2\, T^{(N)}(\mub) = 2 \sum_{i=1}^5 \mbox{Re}
\l[R_{(ii)}^N M^N_i A_{0}(M^N_i) \r]
\ee
from chargino and and neutralino sector,
respectively. The relevant self energies read 
$\Sigma_h^{(C,N)}=\Sigma_h^{(C)}+\Sigma_h^{(N)}$, where 
\be 
16 \pi^2 \Sigma_h^{(C)}(p^2,\mub)= \sum_{i,j=1}^2
\Bigl[ \frac{1}{2}(|L_{ij}^C|^2+|R_{ij}^C|^2)\Bigl(A_{0}(M^C_i)+A_{0}(M^C_j)
 + 
\ee
$$
+((M_i^C)^{2}+(M_j^C)^{2}-p^2)B_0(p^2,M^C_i,M^C_j) \Bigr)+2\mbox{Re} M^C_i M^C_j R_{ij}^C (L_{ij}^C)^* B_{0}(p^2,M^C_i,M^C_j) \Bigr],
$$
\be 
16 \pi^2 \Sigma_h^{(N)}(p^2,\mub)= \sum_{i,j=1}^5
\Bigl[ |R_{(ij)}^N|^2 \Bigl(A_{0}(M^N_i)+A_{0}(M^N_j)+ 
\ee
$$
+((M_i^N)^{ 2}+(M_j^N)^{2}-p^2)B_0(p^2,M^N_i,M^N_j) \Bigr)+2\mbox{Re} M^N_i M^N_j R_{(ij)}^N (R_{(ij)}^N)^* B_{0}(p^2,M^N_i,M^N_j) \Bigr].
$$
The last term in Eq.~(\ref{ChNeutCorr}) is the corrections from the
contribution of chargino and neutralino  into the $W^\pm$ boson
self-energy  
\begin{multline}
 16\pi^2 \, \Pi^{(C,N)}_{WW}(0,\mub) = 
\\ 
 = g^2 \sum_{i=1}^5\sum_{j=1}^2 
\Bigl((C^{L}_{ij}C_{ij}^{L\,*}+C^R_{ij}C_{ij}^{R\,*})\l[ a^2 \l( \ln \frac{a^2}{\mub^2}-\frac{1}{2} \r)+b^2 \l( \ln \frac{b^2}{\mub^2}-\frac{1}{2} \r)+\frac{a^2 b^2}{a^2-b^2}\ln\frac{a^2}{b^2} \r]+
\\
 +2(C^L_{ij}C_{ij}^{R\,*}+C^R_{ij}C_{ij}^{L\,*}) \frac{a b}{a^2-b^2} \l[ - a^2 \l( \ln \frac{a^2}{\mub^2}-1 \r)+b^2 \l( \ln \frac{b^2}{\mub^2}-1 \r)\r]
 \Bigr),
 \label{PiWW}
\end{multline}
where $a=M_{j}^{C}$ and $b=M_{i}^{N}$ are the mass 
eigenstates of  chargino 
and neutralino, respectively. For the explicit 
calculation of Higgs mass (\ref{OneLoopHiggsMass1}), 
one should set $C_{UV}=0$ in 
(\ref{HiggsSPCorr}) and (\ref{ChNeutCorr}).

\subsection{One-loop correction to Yukawa coupling of top quark
\label{TopYukawaAppendix}}
The mass of the Higgs boson at one-loop level is quite sensitive to
the Yukawa  coupling of top quark, $y_t$. Hence, it is important to
include  the RG effects and threshold corrections from top quark
sector for explicit analysis of one-loop corrections to the Higgs
  boson mass in the split NMSSM. Here we  
briefly summarize the results of~\cite{Binger:2004nn} concerning
corrections related to $y_t$. The top quark Yukawa coupling at the
scale $\mub$ can be extracted from its pole mass $M_t=173.2\pm 0.9$
GeV \cite{Lancaster:2011wr},  
\be
y_{t}(\mub)=\sqrt{2}\frac{M_t}{v}(1+\delta_t(\mub)),
\label{Yukawa1} 
\ee 
where the threshold correction $\delta_t(\mub)$ is the sum of the QCD, EW
and split NMSSM terms 
\be
\delta_t(\mub)= \delta_t^{QCD}(\mub) +\delta^{EW}_t(\mub)
+\delta_{t}^{(C,N)}(\mub).
\label{CorrYukawa1} 
\ee
Explicit 3-loop calculation of $\delta_t^{QCD}(\mub)$ was performed
by~\cite{Chetyrkin:1999ys} and at $\mub = M_t$ it yields
\be
\delta^{QCD}_t(\mub=M_t) = 
-\frac{4}{3}\l( \frac{\alpha_3(M_t)}{\pi}\r)  
-9.1 \l( \frac{\alpha_3(M_t)}{\pi}\r)^2-80 \l( \frac{\alpha_3(M_t)}{\pi}\r)^3\approx -0.060.    
\ee
The contribution of the EW term $\delta_{t}^{EW}$ is
negligible~\cite{Binger:2004nn},  $|\delta_{t}^{EW}|< 0.001$. 
The term $\delta_{t}^{(C,N)}$ from chargino and neutralino in split
NMSSM is given through the relation
\be
\delta_t(\mub) = - \frac{\Pi^{(C,N)}_{WW}(0,\mub)}{2 M_W^2}, 
\ee
where $\Pi^{(C,N)}_{WW}(0,\mub)$ is defined by Eq.~(\ref{PiWW}).

\section{Minimization of the effective potential
\label{SectFineTunDim}}

Here we present minimization conitions for the scalar potential of the
model which allow us to express the soft parameters $m^2$,
$\tilde{m}^2$ and $\tilde{m}_N^2$  via the expectation values $v,
v_S$ and $v_P$ (cf. Eq.~(\ref{gener_poten})). To do that, let 
us consider one-loop effective potential at zero temperature  
\be 
V^{eff}_{T=0}=V_{tree}+V^{(1)},
\label{EffPotZeroT}
\ee  
where $V_{tree}$ is the tree level potential, $V_{tree}
\equiv-\mathcal{L}_{V}$, and $V^{(1)}$ is the one-loop  contribution
of fermions, gauge bosons and scalars to the effective potential. 
In the $\overline{\mbox{DR}}$ scheme, $V^{(1)}$ has the form
\cite{Coleman:1973jx} 
\be
V^{(1)}=\sum_i(\pm) \frac{n_i m^4_i}{64 \pi^2}\left( \ln \frac{m_i^2}{q^2}
-\frac{3}{2} \right), \label{1l_eff_pot}
\ee 
here $(+)$ is for bosons and $(-)$ is for fermions and sum runs over all
particles which have field-dependent mass $m_i$ and $n_i$ degrees of
freedom. We choose the renormalization scale $q$ at $100$ GeV.  In
order to define the global minimum of the potential
(\ref{EffPotZeroT}) at the fixed  point $(v,v_S,v_P)$, we expand
$V_{tree}$ in the following  way  
\be
V_{tree}  = -\frac{m^2}{2}v^2 + 
\frac{\tilde{m}^2_N}{2}(v_S^2+v_P^2)+
\frac{\tilde{m}^2}{2}(v_S^2-v_P^2) + V^{>2}_{tree},
\label{Vtree2}
\ee
where $V^{>2}_{tree}$ stands for cubic and quartic terms of 
the tree level potential (\ref{gener_poten}). 
 The vacuum $(v,v_S,v_P)$ at zero 
temperature is determined by the stationary conditions
\be 
\frac{\partial}{\partial v} V^{eff}_{T=0}(v,v_S,v_P) = \frac{\partial 
}{\partial v_S}V^{eff}_{T=0}(v,v_S,v_P) =
\frac{\partial }{\partial v_P}V^{eff}_{T=0}(v,v_S,v_P)=0.  
\label{StationCond}
\ee 
We emphasize that $v$ is fine-tuned to be the vacuum expectation value
of Higgs boson,  $v=246$ GeV. It follows from Eq.~(\ref{Vtree2}) and  
Eq.~(\ref{StationCond}), that squared masses $m^2$, $\tilde{m}^2_N $
and  $\tilde{m}^2$ can be redefined in the following form
\be 
m^2=\frac{1}{v}\frac{\partial}{\partial v}
(V^{>2}_{tree}+V^{(1)}), \quad
 \tilde{m}^2_N=\frac{1}{2}\left(\frac{1}{v_S}\frac{\partial}{\partial v_S}+
\frac{1}{v_P}\frac{\partial}{\partial v_P}\right)
(V^{>2}_{tree}+V^{(1)}),  
\label{msquared1}
\ee 
\be 
\tilde{m}^2=\frac{1}{2}\left(\frac{1}{v_S}\frac{\partial}{\partial v_S}-
\frac{1}{v_P}\frac{\partial}{\partial v_P}\right)
(V^{>2}_{tree}+V^{(1)}).
\label{msquared2}
\ee 
At the tree level this yields
$$
m^2= \frac{1}{2} \tilde{\lambda} v^2 - (\kappa_1-\kappa_2)v_P^2 -
(\kappa_1+\kappa_2)v_S^2, \quad 
\tilde{m}_N^2= \frac{\tilde{A}_k}{\sqrt{2}}\left( 
\frac{v_S}{2} +\frac{v_P^2}{v_S}\right) 
- \frac{\lambda_N}{2}(v_S^2+v_P^2),
$$
$$
\tilde{m}^2=\frac{\tilde{A}_k}{\sqrt{2}}\left(\frac{v_P^2}{v_S}-
\frac{3}{2}v_S \right)-\frac{1}{2}\lambda_N (v_P^2-v_S^2).
$$
Here we neglect contribution arising from  $V^{(1)}$, $\xi$ and 
$\eta$ terms. We stress that in this notation the singlet's VEVs,
$v_S$ and $ v_P$, are the free dimensionful parameters of the split
NMSSM. In our analysis we scan over these two VEVs to find the
parameter space with successful baryogenesis and calculate the
electron and neutron EDMs as obeying the present
experimental constraints.

\end{document}